%% file: main.tex
\documentclass[conference]{IEEEtran}
\usepackage{pifont}
\usepackage{amsmath,amsfonts}
\usepackage{amssymb}
\usepackage{graphicx}
\usepackage{textcomp}
\usepackage{url}
\usepackage{algorithm}
\usepackage[noend]{algpseudocode}
\usepackage{enumerate}
\usepackage{setspace}
\usepackage{amsthm,paralist}
\usepackage{hyperref}
\usepackage[noadjust]{cite} %IEEEtran combine citations
\usepackage{cleveref}
\usepackage{threeparttable}
\usepackage{multirow}
\usepackage{listings}
\usepackage[svgnames]{xcolor}

\hypersetup{
    colorlinks=true,
    linkcolor=ForestGreen,
    citecolor=blue,
    filecolor=magenta,      
    urlcolor=.,
    pdftitle={Overleaf Example},
    pdfpagemode=FullScreen,
    }

\definecolor{verylightgray}{rgb}{.97,.97,.97}

\lstdefinelanguage{Solidity}{
	keywords=[1]{anonymous, assembly, assert, balance, break, call, callcode, case, catch, class, constant, continue, constructor, contract, debugger, default, delegatecall, delete, do, else, emit, event, experimental, export, external, false, finally, for, function, gas, if, implements, import, in, indexed, instanceof, interface, internal, is, length, library, log0, log1, log2, log3, log4, memory, modifier, new, payable, pragma, private, protected, public, pure, push, require, return, returns, revert, selfdestruct, send, solidity, storage, struct, suicide, super, switch, then, this, throw, transfer, true, try, typeof, using, value, view, while, with, addmod, ecrecover, keccak256, mulmod, ripemd160, sha256, sha3}, % generic keywords including crypto operations
	keywordstyle=[1]\color{blue}\bfseries,
	keywords=[2]{address, bool, byte, bytes, bytes1, bytes2, bytes3, bytes4, bytes5, bytes6, bytes7, bytes8, bytes9, bytes10, bytes11, bytes12, bytes13, bytes14, bytes15, bytes16, bytes17, bytes18, bytes19, bytes20, bytes21, bytes22, bytes23, bytes24, bytes25, bytes26, bytes27, bytes28, bytes29, bytes30, bytes31, bytes32, enum, int, int8, int16, int24, int32, int40, int48, int56, int64, int72, int80, int88, int96, int104, int112, int120, int128, int136, int144, int152, int160, int168, int176, int184, int192, int200, int208, int216, int224, int232, int240, int248, int256, mapping, string, uint, uint8, uint16, uint24, uint32, uint40, uint48, uint56, uint64, uint72, uint80, uint88, uint96, uint104, uint112, uint120, uint128, uint136, uint144, uint152, uint160, uint168, uint176, uint184, uint192, uint200, uint208, uint216, uint224, uint232, uint240, uint248, uint256, var, void, ether, finney, szabo, wei, days, hours, minutes, seconds, weeks, years},	% types; money and time units
	keywordstyle=[2]\color{teal}\bfseries,
	keywords=[3]{block, blockhash, coinbase, difficulty, gaslimit, number, timestamp, msg, data, gas, sender, sig, value, now, tx, gasprice, origin},	% environment variables
	keywordstyle=[3]\color{violet}\bfseries,
	identifierstyle=\color{black},
	sensitive=false,
	comment=[l]{//},
	morecomment=[s]{/*}{*/},
	commentstyle=\color{gray}\ttfamily,
	stringstyle=\color{red}\ttfamily,
	morestring=[b]',
	morestring=[b]"
}

\newtheorem{definition}{Definition}
\newtheorem{theorem}{Theorem}
\newtheorem{lemma}[theorem]{Lemma}
\algnewcommand\algorithmicforeach{\textbf{for each}}
\algdef{S}[FOR]{ForEach}[1]{\algorithmicforeach\ #1\ \algorithmicdo}

\newenvironment{myproof}[1]{\noindent \emph{Proof{#1}:}}{\hfill$\square$}

\algnewcommand\algorithmicinput{\textbf{INPUT:}}
\algnewcommand\INPUT{\item[\algorithmicinput]}
\algnewcommand\algorithmicoutput{\textbf{OUTPUT:}}
\algnewcommand\OUTPUT{\item[\algorithmicoutput]}
\algnewcommand\algorithmicassume{\textbf{ASSUME:}}
\algnewcommand\ASSUME{\item[\algorithmicassume]}

\AtBeginDocument{%
  \providecommand\BibTeX{{%
    \normalfont B\kern-0.5em{\scshape i\kern-0.25em b}\kern-0.8em\TeX}}}

\begin{document}

\title{Front-running Attack in Sharded Blockchains and Fair Cross-shard Consensus}

\author{
\IEEEauthorblockN{Jianting Zhang\textsuperscript{$*$}, Wuhui Chen\textsuperscript{$\dag$$\ddag$}, Sifu Luo\textsuperscript{$\dag$}, Tiantian Gong\textsuperscript{$*$}, Zicong Hong\textsuperscript{$\S$}, and Aniket Kate\textsuperscript{$*$$\P$}}
\IEEEauthorblockA{\textsuperscript{$*$}Purdue University,\textsuperscript{$\dag$}Sun Yat-sen University, \textsuperscript{$\ddag$}Pengcheng Laboratory \\
\textsuperscript{$\S$}The Hong Kong Polytechnic University, \textsuperscript{$\P$}Supra Research \\
% Email: \{zhan4674,tg,aniket\}@purdue.edu, \{chenwuh,luosf\}@\{mail,mail2\}.sysu.edu.cn, zicong.hong@connect.polyu.hk
}
\\
% \IEEEauthorblockN{Tiantian Gong}
% \IEEEauthorblockA{Purdue University}
% \and
% \IEEEauthorblockN{Wuhui Chen}
% \IEEEauthorblockA{Sun Yat-sen University / Pengcheng Laboratory}
% \\
% \IEEEauthorblockN{Zicong Hong}
% \IEEEauthorblockA{The Hong Kong Polytechnic University
% }
% \and
% \IEEEauthorblockN{Sifu Luo}
% \IEEEauthorblockA{Sun Yat-sen University
% }
% \\
% \IEEEauthorblockN{Aniket Kate}
% \IEEEauthorblockA{Purdue University / Supra}
}

% \author[1]{Jianting Zhang}
% \author[2,4]{Wuhui Chen \thanks{test}}
% \author[2]{Sifu Luo}
% \author[1]{Tiantian Gong}
% \author[3]{Zicong Hone}
% \author[1,5]{Aniket Kate}
% \affil[1]{Purdue University}
% \affil[2]{Sun Yat-sen University}
% \affil[3]{The Hong Kong Polytechnic University}
% \affil[4]{Pengcheng Laboratory}
% \affil[5]{Supra}

% \IEEEoverridecommandlockouts
% \makeatletter\def\@IEEEpubidpullup{6.5\baselineskip}\makeatother
% \IEEEpubid{\parbox{\columnwidth}{
%     Network and Distributed System Security (NDSS) Symposium 2024\\
%     26 February - 1 March 2024, San Diego, CA, USA\\
%     ISBN 1-891562-93-2\\
%     https://dx.doi.org/10.14722/ndss.2024.23197\\
%     www.ndss-symposium.org
% }
% \hspace{\columnsep}\makebox[\columnwidth]{}}

\maketitle

\input{abstract}

\input{introduction}

\input{preliminaries}

\input{attack}

\input{problemstatement}

\input{overview}

\input{haechidetails}

% \input{security-analysis}
\input{appendics/apdx-proof-v2}

\input{evaluation}

\input{appendics/apdx-relatedworks}

\input{discussion.tex}

\input{conclusion}

% \input{acknowledgement}

% \section*{Acknowledgment}

\bibliographystyle{IEEEtran}
\bibliography{cite}

% \appendix
\appendices
\input{appendix}

\end{document}

%% file: abstract.tex
\begin{abstract} 
%Sharding is a prominent technique for scaling blockchains. A sharded blockchain handles transactions in parallel without introducing inconsistencies by coordinating intra-shard and cross-shard consensus protocols. We observe that the existing sharded systems are vulnerable to transaction ordering manipulations when combining these two protocols. In this work, we formulate and present the first front-running attack against sharded systems with such a finalization fairness issue. This attack allows the adversary to manipulate the execution order of transactions even if the victim's transaction has been processed and packed into a block or even appended to the blockchain by a {\em fair} intra-shard consensus.

Sharding is a prominent technique for scaling blockchains. By dividing the network into smaller components known as shards, a sharded blockchain can process transactions in parallel without introducing inconsistencies through the coordination of intra-shard and cross-shard consensus protocols. However, we observe a critical security issue with sharded systems: transaction ordering manipulations can occur when coordinating intra-shard and cross-shard consensus protocols, leaving the system vulnerable to attack. Specifically, we identify a novel security issue known as finalization fairness, which can be exploited through a front-running attack. This attack allows an attacker to manipulate the execution order of transactions, even if the victim's transaction has already been processed and added to the blockchain by a fair intra-shard consensus. 

%Toward addressing this significant issue, we offer Haechi, a novel cross-shard protocol immune to this attack. The key idea is to introduce an ordering phase between transaction processing and execution, by which the execution order of transactions is the same as the processing order, ensuring the so-called finalization fairness. Considering different shards handle transactions at various consensus speeds, Haechi adopts a finalization fairness algorithm to achieve a globally fair order with minimal performance loss caused by sluggish shards. In addition, a global order provides a strong consistency among shards, enabling Haechi to achieve better parallelism in handling conflicting transactions across shards. This makes Haechi promising for supporting popular smart contracts in the real world. We implement Haechi based on Tendermint and perform an extensive evaluation on a geo-distributed AWS environment. The experimental results demonstrate that, compared to the existing cross-shard consensus protocols, Haechi sacrifices very little in terms of performance while delivering finalization fairness. 

To address the issue, we offer Haechi, a novel cross-shard protocol that is immune to front-running attacks. Haechi introduces an ordering phase between transaction processing and execution, ensuring that the execution order of transactions is the same as the processing order and achieving finalization fairness. To accommodate different consensus speeds among shards, Haechi incorporates a finalization fairness algorithm to achieve a globally fair order with minimal performance loss. By providing a global order, Haechi ensures strong consistency among shards, enabling better parallelism in handling conflicting transactions across shards. These features make Haechi a promising solution for supporting popular smart contracts in the real world. To evaluate Haechi's performance and effectiveness in preventing the attack, we implemented the protocol using Tendermint and conducted extensive experiments on a geo-distributed AWS environment. Our results demonstrate that Haechi can effectively prevent the presented front-running attack with little performance sacrifice compared to existing cross-shard consensus protocols.
\end{abstract}

%% file: introduction.tex
\section{Introduction}\label{section: introduction}
Sharding is a prominent approach to enhance blockchain scalability and has been researched extensively by academia \cite{Elastico,Omniledger, Rapidchain,Monoxide,gearbox,free2shard,krol2021shard} and deployed in the industry \cite{ethereum2,harmony,zilliqa,elrond,nightshade,ontology}. Its main idea is to horizontally partition the entire blockchain into multiple self-maintaining shards. With more nodes participating in consensus, sharding can assign more shards to process transactions in parallel, bringing higher throughput. However, there is no free lunch here: along with an intra-shard consensus protocol that can be instantiated with any standard Byzantine Fault Tolerant (BFT) state machine replication (SMR) protocols \cite{tendermint-bft, hotstuff}, sharded blockchains also need a cross-shard consensus protocol that is specifically designed to handle transactions involving entities and data from two or more shards. The key to a cross-shard consensus protocol is to ensure the isolation property that there is no inconsistent access to the same data and the atomicity property that all relevant shards either commit or abort the cross-shard transaction.

\begin{figure}
    \centering
    \includegraphics[width=3.5in]{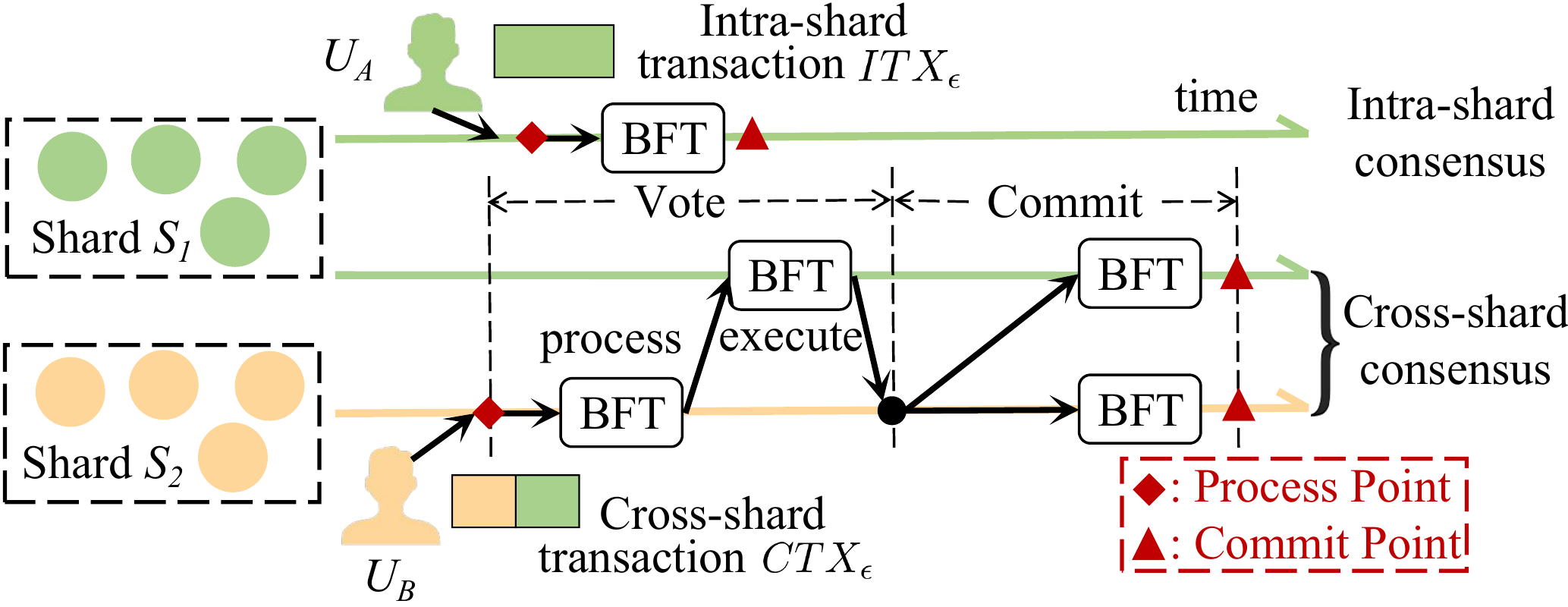}
    \caption{Finalization unfairness: Assume shard $S_1$ maintains a contract $\epsilon$. The intra-shard transaction $ITX_{\epsilon}$ is executed and committed before the cross-shard transaction $CTX_{\epsilon}$ even though $CTX_{\epsilon}$ is processed earlier than $ITX_{\epsilon}$.}
    \label{fig:motivation}
    % \vspace{-4mm}
\end{figure}

Most previous sharded systems adopt a two-phase (2P) protocol to handle cross-shard transactions, motivated by the two-phase commit protocol of traditional distributed databases \cite{2PC}. The 2P cross-shard protocol works as follows: First, in the vote phase, all shards relevant to the cross-shard transaction modify and lock local states by running one instance of the intra-shard consensus; then, in the commit phase, another instance of intra-shard consensus is used to commit the transaction. For example, as shown in Fig. \ref{fig:motivation}, a user $U_B$ in shard $S_2$ constructs a cross-shard contract transaction $CTX_{\epsilon}$ to call a smart contract $\epsilon$ maintained by shard $S_1$. The sender's shard (i.e., $S_2$) first {\em processes} $CTX_{\epsilon}$, e.g., deducting $U_B$'s balance used for this transaction. Then, the contract's shard (i.e., $S_1$) {\em executes} the contract $\epsilon$ with the data generated by $S_2$ and modifies $\epsilon$'s state data. $CTX_{\epsilon}$ will be committed if these two data modifications are operated successfully. 

An observation is that transaction \emph{processing} and contract \emph{execution} for a cross-shard contract transaction are completed by different shards and introduce multiple instances of intra-shard consensus. \emph{A cross-shard transaction will thus be packed into many blocks from multiple shards, involving multiple transaction orders}. This is inevitable due to the separation of data in a sharded system. In contrast, these operations for an intra-shard transaction are completed by only one shard via one instance of intra-shard consensus. Therefore, there exists a \emph{processing-execution difference} between intra-shard and cross-shard contract transactions. Such a difference is not an issue if the called contract is agnostic to transaction order, e.g., a transfer contract that previous sharded systems focus on \cite{ahl, byshard, sharper, COSPLIT}. Unfortunately, there are also contracts where the transaction order is crucial, such as decentralized finance (DeFi) \cite{sandwich-attack}, initial coin offerings \cite{eskandari2019sok}, and gambling \cite{torres2021frontrunner}.
For instance, recently Daian et al.\ \cite{flashboys} and Qin et al.\ \cite{quantify-front-attack} show that the adversarial manipulation of transaction order in Ethereum DeFi can extract millions of USD from users.
In this work, we call such contracts \emph{order-sensitive contracts (OSCs)} and the problem relevant to the processing-execution difference when calling OSCs as the finalization fairness problem. Fairness is a loaded term in the blockchain field, and we formally define finalization fairness for a blockchain sharded system in \S~\ref{section3-fairness-definition}. Informally, it means that transactions\footnote{In the rest of the paper, transactions represent contract transactions, and transaction executions represent executions of the called contracts.} are executed and committed in the order they are processed.

We formulate and present the \emph{front-running attack} triggered by this fairness problem (\S~\ref{section-attack-details}). To the best of our knowledge, this is the first front-running attack in sharded systems. This attack allows the adversary to manipulate the transaction order to have their transactions executed and finalized before the victims' transactions. For instance, an attacker strategically constructs a transaction (e.g., an intra-shard transaction $ITX_{\epsilon}$) to target a cross-shard transaction (e.g., $CTX_{\epsilon}$) created by the victim. Since the processing and execution of cross-shard transactions are separated, the attacker's transaction $ITX_{\epsilon}$ can be inserted between the processing and execution of the victim's transaction $CTX_{\epsilon}$. This will lead that $ITX_{\epsilon}$ is finally executed before $CTX_{\epsilon}$ even though $CTX_{\epsilon}$ is processed and stored on the blockchain earlier than $ITX_{\epsilon}$ (as shown in Fig. \ref{fig:motivation}). Such a front-running attack is inexpensive without transaction fee bidding contests \cite{flashboys, sandwich-attack}, but more likely to succeed in sharded systems because of the existence of the processing-execution difference. As we discuss in \S~\ref{section-attack-impact}, the front-running attack can not only threaten the security of the above OSCs, but also lead to unbalanced shards, eventually sacrificing the scalability of sharded systems.

To prevent this front-running attack in sharded systems, we propose Haechi\footnote{Haechi is a mythical creature representing justice and fairness.}, a novel cross-shard protocol (\S~\ref{sec-haechi-overview}). Haechi is a \emph{process-order-execute-commit} (POEC) consensus model whose key idea is to introduce a globally fair ordering phase between transaction processing and execution phases. With the ordering phase, Haechi ensures that the execution order is consistent with the processing order of OSC transactions, thus providing the finalization fairness for sharded systems. In addition to satisfying the finalization fairness, a globally fair order achieves another desirable feature for a sharded system. Specifically, since a global order ensures the consistency of transaction commitment, Haechi supports non-blocking processing for transactions, i.e., conflicting transactions that call the same contracts can be handled continuously in parallel even though some of them are not committed yet. Since in a practical scenario, there are some popular smart contracts that are called by users frequently, the feature of non-blocking processing makes Haechi more practical and promising in supporting real-world smart contracts. 

However, since shards have various consensus speeds in a sharded system, establishing a globally fair order for a sharded system faces new challenges. \emph{The challenges come from the in-flight transactions.} Specifically, due to the various consensus speeds, committing a block in the processing phase takes longer for a sluggish shard than it does for a prompt shard. As a result, when transactions in the prompt block are ordered, those in the sluggish block are still being processed and will skip the ordering phase. These \emph{in-flight transactions} of the sluggish block, even if being processed earlier, are ordered after the transactions of the prompt block, compromising finalization fairness. To eliminate in-flight transactions, a straightforward solution is to \emph{synchronize} all shards block-by-block, i.e., any shard is not allowed to propose new blocks until all shards complete consensus for their current blocks, by which all shards have the same block heights and thus there are no in-flight transactions on any block height. However, such a synchronization mechanism is low-efficiency and far from the practicality of sharded systems.

To solve the challenges, Haechi adopts a \emph{finalization fairness algorithm} to coordinate a globally fair order for a sharded system. Instead of asking all shards to synchronize their block heights, the algorithm allows each shard to work independently without being affected by the ordering phase or other shards, while yet guaranteeing finalization fairness. To achieve this, Haechi first uses the \emph{block timestamp} as the ordering indicator. Since the block timestamp can represent the time when transactions are processed and is incremental in each shard, it can be used to establish a \emph{deterministic global order} in which transactions that are processed first must be ordered first. Furthermore, since the block timestamp is publicly verifiable and immutable once a block is committed, Haechi prevents the adversary from manipulating the order by forging the block timestamp. Then, considering in-flight transactions, Haechi adopts a so-called \emph{at-least-one} ordering rule. Specifically, we observe that our protocol can catch enough information and start an ordering phase without waiting to receive all in-flight transactions as long as it receives one block from each shard. The finalization fairness algorithm allows Haechi to reduce performance loss caused by the global ordering phase. 
We present security analysis in \S~\ref{sec-security-proof-detail}.
% To solve the challenges, Haechi first assigns a dependent \emph{beacon chain\footnote{In the previous sharded systems, a beacon chain (or called identity chain \cite{Omniledger}, reference chain \cite{Rapidchain, ahl, rivet}, etc.) is widely used for global configuration maintenance \cite{Omniledger, Rapidchain, harmony, eth2-beaconchain} or cross-shard coordination \cite{ahl, rivet, gearbox}.}} to coordinate the ordering phase and uses the \emph{block timestamp} as the ordering indicator, making our protocol easy to be integrated into current sharded systems. Since the block timestamp is publicly verifiable and immutable once a block is committed, nodes in the beacon chain can establish a \emph{deterministic global order} for transactions of blocks received from shards. Then, we design a \emph{finalization fairness algorithm}. This algorithm enables any transaction to be eventually ordered by the beacon chain fairly without synchronization between the beacon chain and shard chains. In other words, transactions are guaranteed to be ordered only based on the timestamp of the block that they are processed in, but not the network delay or the consensus speed of shards. Since the block timestamp indicates the time when transactions are processed, such a guarantee satisfies the finalization fairness where transactions that are processed first should be executed first.

Overall, this work makes the following contributions:
\begin{itemize}[-]
    \item We formulate and present the front-running attack against the current 2P cross-shard consensus protocol, which, to the best of our knowledge, is the first front-running attack in sharded systems. This front-running attack can be devastating: (i) shards receiving unbalancing transaction requests, sacrificing the scalability of sharded systems; (ii) OSCs built on sharded systems pose a significant security risk. These unfavorable outcomes further limit applications of sharding technology.
    % This front-running attack can lead that shards receiving unbalancing transaction requests, eventually sacrificing the scalability of sharded systems. 
    \item We give a new fairness definition for sharded systems, called finalization fairness, which demands that any two transactions calling the same OSC must be executed in the same order as they are processed. It is a requisite for preventing front-running in sharded systems. 
    \item We design Haechi, a novel cross-shard protocol with two improvements: (i) it provides finalization fairness and prevents the front-running attack in a sharded system; (ii) it enhances the efficiency of handling OSC transactions. 
    % We also give complete security proof in \cref{sec-security-analysis-detail}.
    \item We implement Haechi with Tendermint \cite{tendermint-bft} and compare it with prior state-of-the-art cross-shard consensus protocols. We evaluate them by running up to 990 nodes on geo-distributed AWS EC2 instances spread across ten regions, showing that Haechi can achieve 13,000+ TPS, less than 6s intra-shard confirmation latency, and less than 17s cross-shard confirmation latency with 33 shards. 
    We also evaluate the presented cross-shard front-running attacks for different cross-shard protocols, showing that Haechi can effectively prevent such attacks while existing 2P cross-shard protocols are vulnerable to them.
\end{itemize}

%% file: preliminaries.tex
\section{Preliminaries}\label{section-preliminaries}
In this section, we start with an example of order-sensitive contracts (OSCs) in a blockchain. Then, we give a concrete working process on how the classic 2P cross-shard consensus protocol  handles a cross-shard OSC transaction.
\subsection{An Example of Order-sensitive Contract}\label{section-preliminaries-example}
OSCs are a type of smart contract whose states are relevant to transactions order. For instance, if transactions $TX_1$ and $TX_2$ call an OSC, executing $TX_1, TX_2$ and executing $TX_2, TX_1$ will result in different OSC states. For illustration purposes, we give an example of OSC that is derived and simplified from Uniswap \cite{uniswap}, one popular decentralized exchange contract. This OSC allows users to buy \emph{tokens} with their \emph{coins}, where tokens are the data of contract state and coins are the data of user accounts\footnote{See differences between tokens and coins on https://blog.liquid.com/coin-vs-token.}. 

\noindent\textbf{TokensExchange Contract.} Fig. \ref{code-dex-contract} presents our example, TokensExchange, written in Solidity \cite{solidity}. TokensExchange consists of: (i) a state variable \emph{factory\footnote{For illustration, we assume the token pools (i.e., UniswapPair contracts) are a part of the state of this Router contract, recorded in the variable \emph{factory}.}} that records core information of the current exchanging market, like tokens pairs, price, and liquidity (Line 3); (ii) a contract function \emph{swapCoinForTokens} realized to swap a specific token with the user's coins (Line 5). This function has three parameters (i.e., input data). The parameter \emph{path} specifies multiple intermediate Token addresses used for this token exchange. The parameter \emph{to} represents the address of the Token that the user is swapping. The parameter \emph{msg.value} represents the number of coins that the user pays for this exchange. The whole token exchanging process is as follows. First, the contract computes how many actual tokens the user can get with the input coins, i.e., computes the token price (Line 8). Then, the contract transfers the coins into a wrapper token existing in the \emph{factory} (Line 9). For example, in Ethereum, the coin \emph{Ether} will be wrapped into a type token in the \emph{factory}, called \emph{WETH}. This operation is used to provide a more secure and convenient way for token exchanges. Finally, in Lines 10-15, the exchanging market updates the price and liquidity of the involved tokens in the routing path, and the user receives the fitting amount of tokens for its exchanged coins. 

According to the function realized in TokensExchange, if a contract transaction is committed successfully, then the user will lose its coins for obtaining relevant tokens, and the contract will update its contract state.

\begin{figure}[t]
	\setlength\abovecaptionskip{-0.1\baselineskip}
	\setlength\belowcaptionskip{-1.5\baselineskip}   % 
	\centering
\begin{lstlisting}[language=Solidity, linewidth={1.0\linewidth}]
contract TokensExchange {
	// state variables of the contract
	address public factory;
	// method to exchange a token with coin
	function swapCoinForTokens(address[] calldata path, address to, uint msg.value)
	{
		// msg.value is the user's coins
		amounts = getAmountsOut(msg.value, path);
		transferCoin(factory, amounts[0]);
		for (uint i; i < path.length - 2; i++ {
			address _to = factory.getTokensAdd(path[i+1]);
			swapAndUpdate(factory, path[i], path[i+1], amount[i+1], _to);
		}
		swapAndUpdate(factory, path[path.length-2], path[path.length-1], amount[path.length-2], to);
	}  
	// other state variables and methods...
}
\end{lstlisting}
	\caption{The code of a simple decentralized exchange contract.}
	\label{code-dex-contract}
 % \vspace{-5mm}
\end{figure}

\subsection{2P Cross-shard Protocol in Sharded Systems}\label{section-preliminaries-2P}
Now, consider the implementation of such an OSC in a sharded system. We assume a user in shard $S_1$ constructs a cross-shard transaction $CTX_\epsilon$ to call TokensExchange contract (denoted as $\epsilon$) maintained by shard $S_2$. In this example, $CTX_\epsilon$ will modify the data of two accounts in two operations. The first, i.e., \emph{transaction processing}, is that $CTX_\epsilon$ withdraws coins (i.e., the \emph{msg.value}) of the user's account in $S_1$. The second, i.e., \emph{contract execution}, is that $CTX_\epsilon$ modifies the contract state of $\epsilon$ (i.e., the state variable \emph{factory}) in $S_2$, including transferring tokens to the user, updating token price and liquidity. We emphasize that in a scenario of calling contracts, the later operations are invoked only after the former operations are completed. Because the later operations rely on data generated after the former operations and contract interactions are unpredictable \cite{forerunner, PROPHET}, e.g., a contract invokes other different contracts based on the block height when it is executed, but the block height is uncertain.

\begin{figure}
    \centering
    \includegraphics[width=3.0in]{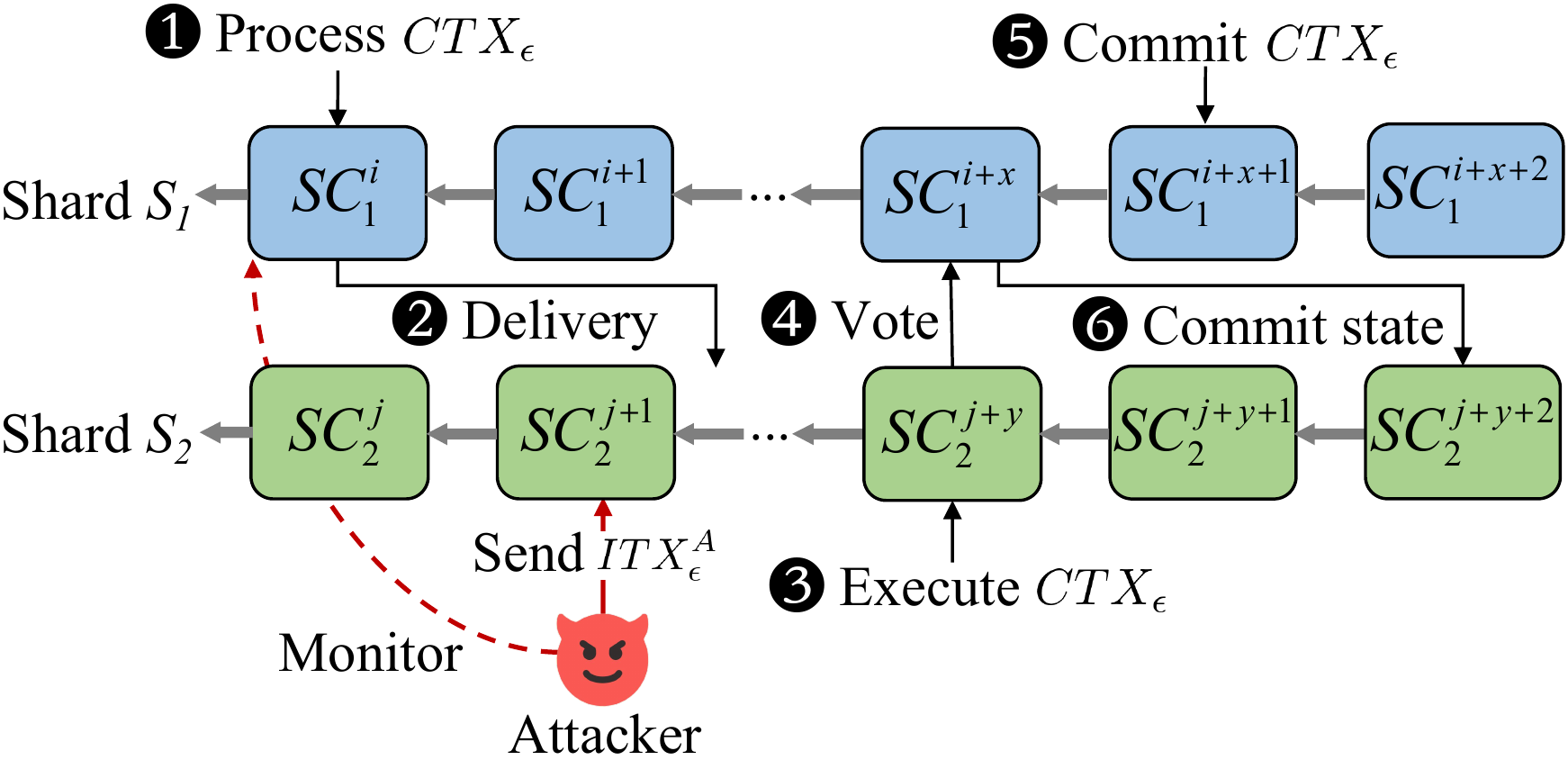}
    \caption{The process of 2P cross-shard consensus protocol with a shard-based coordinator.}
    \label{fig:2p-example}
    % \vspace{-5mm}
\end{figure}

Fig. \ref{fig:2p-example} shows the process of handling $CTX_\epsilon$ with the 2P cross-shard protocol in a sharded system. Similar to traditional databases, the 2P cross-shard protocol is also driven by a coordinator, such as a client \cite{Omniledger,androulaki2018channels}, or a shard \cite{chainspace,Rapidchain, ahl, byshard}. In this example, we assume the coordinator is the shard that manages the account of the transaction's sender, i.e., $S_1$. The 2P cross-shard consensus protocol contains two phases: the vote phase and the commit phase.

\noindent\textbf{Phase 1: Vote for the processing and execution results.} In this vote phase, the coordinator $S_1$ leads messages delivery among all related shards for collaboratively handling $CTX_{\epsilon}$. The specific process can be concluded as follows:
\begin{asparadesc}
    \item [Step \ding{182}:] $S_1$ processes $CTX_{\epsilon}$, where it will modify the data of the transaction's sender (i.e., the user's coins). Specifically, $S_1$ checks if the user has sufficient coins to pay for the transaction (i.e., \emph{msg.value} used for exchange), and locally withdraws these coins from the user's account. Then, $CTX_{\epsilon}$ is packed into a new block $SC_1^i$ by a round of intra-shard consensus, and the relevant coins of the user will be locked for paying until the cross-shard consensus is completed.
    \item [Step \ding{183}:] $S_1$ creates a message used for contract execution. Specifically, $S_1$ packs the input data (i.e., function parameters \emph{path, to}, and \emph{msg.value} in this example), the $\epsilon$'s address, and a commitment quorum certificate on the block $SC_1^i$, into a relay message. This message is used as a proof and request showing that a cross-shard transaction has been processed successfully by the sender's shard.
    Finally, the relay message is forwarded to $S_2$ for the contract execution.
    \item [Step \ding{184}:] The contract shard $S_2$ executes $CTX_{\epsilon}$. When receiving the relay message after a network delay (e.g., after $y$ blocks), $S_2$ verifies its validity. Then, $S_2$ retrieves $\epsilon$ according to its address and executes it with the input data by a round of intra-shard consensus. After that, $S_2$ locally modifies and locks the changing states of $\epsilon$ (i.e., \emph{factory} in this example). The relay message and changing states are recorded into $SC_2^{j+y}$.
    \item [Step \ding{185}:] $S_2$ votes for the commitment. If the execution is successful, $S_2$ sends the execution results with a \emph{vote-for-acceptance} to $S_1$. Otherwise, it sends a \emph{vote-for-abort} to $S_1$.
\end{asparadesc}

\noindent\textbf{Phase 2: Commit the transaction and state changes.} In the second commit phase, the coordinator $S_1$ collects all voting results to determine if $CTX_{\epsilon}$ can be committed.
\begin{asparadesc}
    \item [Step \ding{186}-\ding{187}:] If $S_1$ receives a vote-for-acceptance from $S_2$, $S_1$ unlocks the relevant coins of the user and commits $CTX_{\epsilon}$ into its ledger $SC_1$ (Step \ding{186}). At the same time, the coordinator $S_1$ sends a \emph{commit-for-acceptance} to $S_2$ where $S_2$ will unlock and commit the changing contract states of $\epsilon$ (Step \ding{187}). In contrast, $S_1$ aborts the commitment of $CTX_{\epsilon}$ and sends a \emph{commit-for-abort} to $S_2$ for aborting state changes of $\epsilon$.
\end{asparadesc}

\noindent\textbf{The processing-execution difference.} In this example, we find that handling a cross-shard contract transaction is completed by multiple shards. In particular, transaction processing operation, i.e. Step \ding{182}, is completed by $S_1$ while transaction execution operation, i.e., Step \ding{184}, is completed by $S_2$. In contrast, handling an intra-shard contract transaction only involves one shard. For example, assume a user $U_B$ in $S_2$ calls $\epsilon$ by constructing an intra-shard transaction $ITX^B_{\epsilon}$. Since $S_2$ maintains both $U_B$'s account and $\epsilon$'s account, $ITX^B_{\epsilon}$ can be committed by $S_2$ independently via only one intra-shard consensus. Therefore, there exists a processing-execution difference between intra-shard and cross-shard transactions. In the next section, we will present how the attacker utilizes such a processing-execution difference to launch a front-running attack.

%% file: attack.tex
\section{Front-running in Sharded Systems}\label{section-attack-details}
% In this section, we will present the front-running attack based on the example in \cref{section-preliminaries}, followed by its impacts on a sharded system.  

\subsection{Attacking Process}\label{section-attack-process}
Recall from the process of 2P cross-shard consensus in Fig. \ref{fig:2p-example}, the called contract $\epsilon$ starts to be executed with $CTX_\epsilon$ only after $CTX_\epsilon$ is processed by $S_1$ via an intra-shard consensus and is forwarded to $S_2$ after a period of network latency. Obviously, \emph{there is a time difference between the time when a cross-shard transaction is processed and the time when the called smart contract is actually executed}.

\begin{figure}
    \centering
    \includegraphics[width=2.8in]{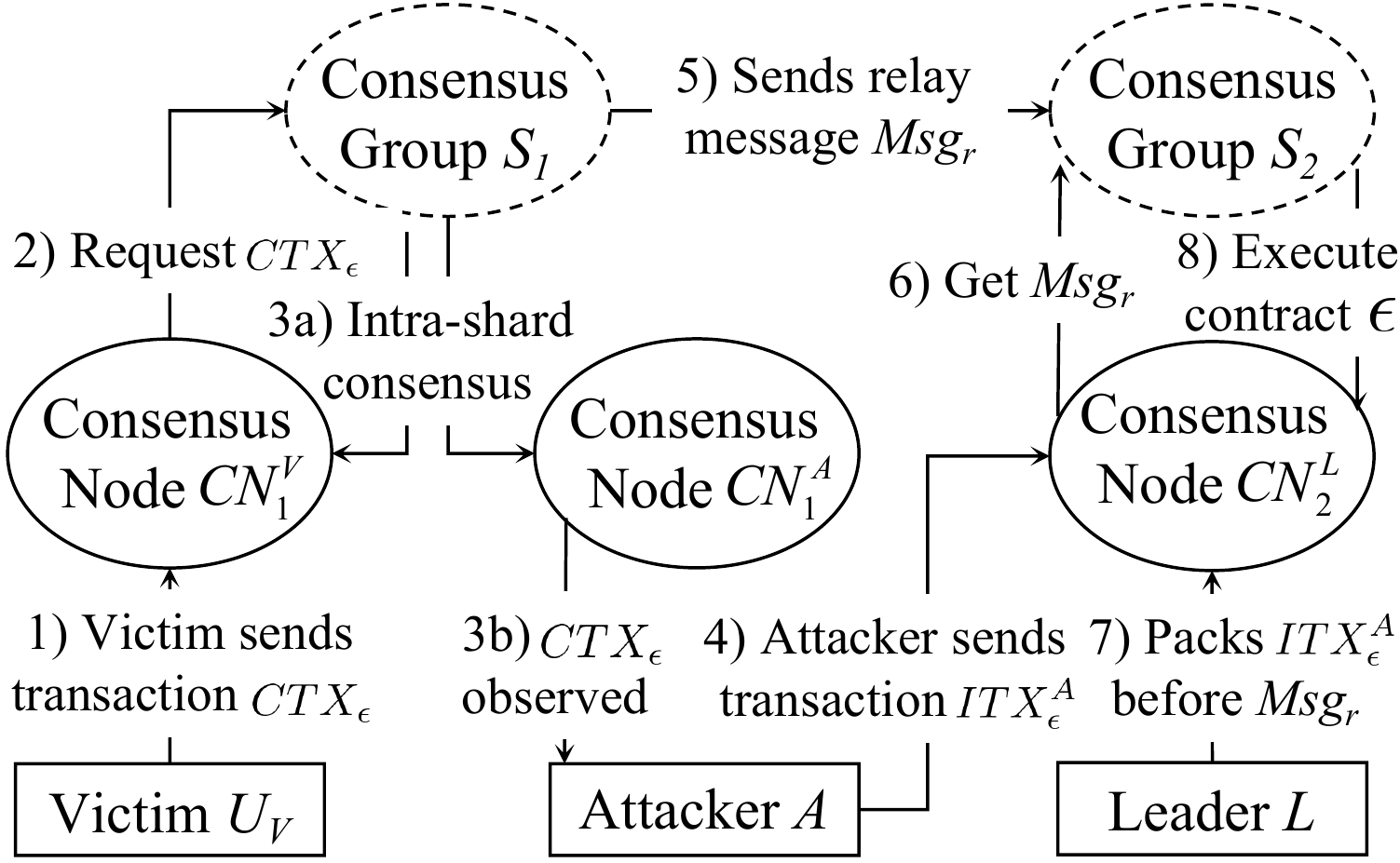}
    \caption{Intra-shard attacking: the attacker monitors the target shard chain and requests a front-running transaction to be executed before the victim's cross-shard transaction.}
    \label{fig:attack-details}
    % \vspace{-5mm}
\end{figure}

\noindent \textbf{Intra-shard attacking model.} The attacker can utilize such a time difference to launch a front-running attack against the cross-shard transaction with an intra-shard transaction. Briefly, as transaction processing and execution are separated for a cross-shard transaction, the attacker constructs and has its intra-shard transactions executed between the processing and execution of the victim's cross-shard transaction. Fig. \ref{fig:attack-details} describes the attacking process concretely, and a brief illustration is shown at the bottom of Fig. \ref{fig:2p-example}. Specifically, to attack $CTX_{\epsilon}$, the attacker $A$ first monitors the network of $S_1$, e.g., it runs a consensus node $CN_1^A$ in $S_1$. Once the attacker observes $CTX_{\epsilon}$ will appear in the next new block, it creates a front-running intra-shard transaction $ITX^A_{\epsilon}$ with its account managed by $S_2$ and sends $ITX^A_{\epsilon}$ to the shard leader $CN_2^L$ of $S_2$. Note that the attacker can observe $CTX_{\epsilon}$ in advance rather than waiting for the new block to be finally appended into the shard chain $SC_1$. That is because a BFT-based intra-shard consensus always requires several stages before reaching a final agreement, e.g., PBFT \cite{pbft} contains pre-prepare, prepare, and commit stages, of which the attacker can receive the new block at the pre-prepare stage. Furthermore, since $CTX_{\epsilon}$ is being processed by $S_1$ (Step \ding{182}) and requires a period of transmission time before arriving at $S_2$ (Step \ding{183}), $ITX^A_{\epsilon}$ is more likely to be executed by $S_2$ and changes the contract state of $\epsilon$ before $CTX_{\epsilon}$ (i.e., before Step \ding{184}). Thus, although $CTX_{\epsilon}$ is processed by the sharded system earlier than $ITX^A_{\epsilon}$, $ITX^A_{\epsilon}$ is finally executed and committed earlier than $CTX_{\epsilon}$. Finally, the victim $U_V$ is front-running attacked by the attacker.

\noindent \textbf{More flexible attacking model.} The above attacking model is based on the processing-execution difference between intra-shard and cross-shard transactions. This requires the attacker to create multiple accounts in different shards if it wants to front-run multiple contracts from different shards with its intra-shard transactions. In addition, the attacker can also adopt a more flexible attacking model, called \emph{cross-shard attacking}, by utilizing the processing-execution difference between cross-shard transactions from distinct shards (see appendix \ref{appendix-general-attack} for more details). The idea is that the attacker only creates accounts in a shard with the fastest consensus speed. In this case, the attacker can still front-run the victims' transactions with its cross-shard transactions. Specifically, the shard where the attacker's accounts are located processes transactions faster than other shards, allowing the attacker's cross-shard transactions to reach the target contract's shard before the victims' cross-shard transactions. As a result, the target contract's shard potentially executes the attacker's transactions before the victims' transactions, resulting in successful front-running attacks.

\subsection{Impacts on Sharded Systems} \label{section-attack-impact}
We note that to launch such a front-running attack in a sharded system, the adversary only needs to monitor the network and register accounts in some shards. It neither needs to compromise the shard security nor involves transaction fee bidding contests. Such low-cost attacks will allow the adversary to conduct front-running transactions intensively, bringing devastating impacts to a sharded system. 

On the one hand, the front-running attack will compromise the scalability of a sharded system. Specifically, to reduce the risk of being front-run, a user would like to register an account in the contract's shard, by which it can avoid the 2P cross-shard consensus. However, such a strategy leads to workload unbalancing among shards, i.e., distinct shards will manage a various number of accounts and handle a various number of transactions. In a real-world scenario, some popular contracts occupy the major transaction volume of the network. For instance, according to the data provided by Etherscan\footnote{https://etherscan.io/stat/dextracker?range=30}, Uniswap accounts for nearly $11\%$ of daily transactions in the Ethereum network from Nov. 25th, 2022, to Dec. 25th, 2022. 
% (2,409,812 + 1,063,114) / 31,574,643
The unbalancing workload among different shards will significantly limit the scalability of sharding technology, i.e., even as the number of shards increases, a sharded system cannot handle more transactions because most transactions are created in only several specific shards. On the other hand, the front-running attack also threatens the security of many decentralized applications that are affected by transaction ordering manipulations significantly, such as DeFi and gambling.

% \textbf{(2) Loss of users asset.} Another impact of the front-running attack is that DEXs users will obtain fewer tokens than expected in a sharded system. For example, we assume a victim user plans to buy 200 UNI tokens with 1 Ether coin based on the latest price of UNI (i.e., 0.005 Ether per UNI) from the contract TokensExchange. However, an attacker conducts a front-running transaction that also trades 1 Ether for 200 UNI. According to the rule of price variation in DEXs, buying the token could increase its token price, e.g., the price of UNI changes to 0.01 after TokensExchange executes the front-running transaction. Therefore, by the time TokensExchange executes the victim transaction, the victim user can only obtain 100 UNI, which is half of what it could get before. 
% Front-running attacks have generated significant impacts on real-world DEXs in Ethereum. According to work \cite{quantify-front-attack}, a type of front-running attack called sandwich attack has yielded 1.51M USD in profit within two years. The front-running attack may become more serious in a sharded system due to the low-cost and high-success attacking behaviors.

It is crucial to prevent such front-running attacks so that a sharded system can be applied to a practical scenario supporting thousands of decentralized applications. However, to the best of our knowledge, none of the current sharded systems consider such issues. This paper thus proposes a novel cross-shard consensus protocol, called $Haechi$, to fill this gap. 

%% file: problemstatement.tex
\section{Problem Statement}\label{section-problem-statement}
%Before presenting Haechi, in this section, we give related definitions, assumptions, and our goals for this work.

\subsection{System Definitions}\label{sec-network-model}
We target the setting of state-sharded systems with smart contract capabilities, where the whole ledger is partitioned among several sets of nodes by a Sybil-resistance protocol, e.g., Proof-of-Work (as in \cite{Elastico, Omniledger, Rapidchain}) or Proof-of-Stake (as in \cite{ethereum2, ontology, elrond}). Besides, shards adopt a BFT-based protocol \cite{pbft, hotstuff, tendermint-bft} as their intra-shard consensus. There are $(m+1)$ shards in the system, in which a shard $S_i(1 \leq i \leq m)$ maintains a separate shard chain $SC_i=\{SC_i^1, SC_i^2, ...\}$, where $SC_i^b$ is the $b$-th block in $SC_i$. Each shard block is attached with a block timestamp to represent the time of block generation. Furthermore, there is a beacon shard $S_0$ that is responsible for maintaining a \emph{beacon chain} $BC=\{BC^1, BC^2, ...\}$. The beacon chain\footnote{In the previous sharded systems, the beacon chain also called the identity chain \cite{Omniledger} or reference chain \cite{Rapidchain, ahl, rivet}, etc.} is widely used by previous sharded systems, which is used to coordinate cross-shard communications \cite{ahl, elrond, rivet, gearbox} or maintain the global configuration for the sharded system \cite{Omniledger, Rapidchain, harmony, eth2-beaconchain}, such as shard number, node identity, epoch randomness, etc.

\subsection{Trust Assumptions and Adversary Model}\label{section3-trust-model}
We assume there is a reliable and secure shard reconfiguration (or resharding) mechanism like prior sharded systems \cite{Omniledger, Rapidchain}, by which the sharded system adjusts all shards periodically to ensure that the honest nodes always control each shard. 
Specifically, a shard $S_i(0 \leq i \leq m)$ always satisfies $3f_i+1 < |S_i|$, where $f_i$ is the number of byzantine nodes and $|S_i|$ is the shard size. In light of this assumption, the following security properties are always guaranteed by the intra-shard BFT protocol in each shard:
\begin{itemize}
    \item Intra-shard safety. Any two honest nodes of the same shard store the same prefix ledger.
    \item Intra-shard liveness. If a transaction is received by at least one honest node, then
    the transaction will be eventually handled and get a response from the shard.
\end{itemize}

Furthermore, we assume that the adversary can behave arbitrarily to launch attacks to have its transactions executed before the victim's transactions. 
However, the adversary is computationally bounded.
% and cryptographic primitives are secure.

%We assume a partially synchronous network model, where there exists an unknown finite time bound $\Delta$ such that all messages sent by honest nodes can be delivered within $\Delta$. 

We assume a partially synchronous network model since each shard independently runs a partially synchronous SMR protocol (i.e., Tendermint BFT~\cite{tendermint-bft} in this work). However, we emphasize that Haechi can also work in an asynchronous network since it does not rely on any time-bound. Specifically, if shards run an asynchronous SMR protocol (e.g., Tusk~\cite{danezis2022narwhal}), then the network assumption could be asynchrony.

% We plan to develop our inter-shard communication protocol to operate completely asynchronously (i.e., the adversary can delay a message between two honest parties by any finite amount of time as long as the message must eventually be delivered.) As a result, the network/communication model of the employed intra-shard SMR protocol determines our communication model assumption. To illustrate, if each shard independently runs a partially synchronous BFT protocol (such as Tendermint BFT~\cite{tendermint-bft} used in this work), then the system would require the standard partially synchrony assumption~\cite{DLT}.

\subsection{Fairness in Sharded Systems}\label{section3-fairness-definition}
The front-running attack discussed in \S~\ref{section-attack-details} reflects a fairness issue where the adversary can manipulate the execution order of transactions independent of their processing order. Therefore, to give a clear definition of fairness for sharded systems, we first define two ordering relations for transactions: processing order and execution order.

\begin{definition}[Processing order]\label{definition-processing-order}
    Let $\prec_\mathbf{P}$ denote a processing order relation in a sharded system, $TX_{\mapsto \epsilon} \in SC_i^j$ denote that transaction $TX_{\mapsto \epsilon}$ calling contract $\epsilon$ is recorded in block $SC_i^j$. Given transactions $TX^1_{\mapsto \epsilon}$ and $TX^2_{\mapsto \epsilon}$, we define $TX^1_{\mapsto \epsilon} \prec_\mathbf{P} TX^2_{\mapsto \epsilon}$ if one of the following three conditions is satisfied: 
    \begin{compactenum}[(i)]
        \item $TX^1_{\mapsto \epsilon}, TX^2_{\mapsto \epsilon} \in SC_i^m$, and $f_{idx}(TX^1_{\mapsto \epsilon}) < f_{idx}(TX^2_{\mapsto \epsilon})$, where $f_{idx}(\cdot)$ is a function returning the indexes of transactions in a block.
        \item $TX^1_{\mapsto \epsilon} \in SC_i^m$, $TX^2_{\mapsto \epsilon} \in SC_i^n$, and $m<n$, where $m$ and $n$ represent the block height of the same chain $SC_i$.
        \item $TX^1_{\mapsto \epsilon} \in SC_i^m$, $TX^2_{\mapsto \epsilon} \in SC_j^n$, and $f_{bt}(SC_i^m) < f_{bt}(SC_j^n)$, where $f_{bt}(\cdot)$ is a function returning the timestamp of a block.
    \end{compactenum}
\end{definition}
In a sharded system, two transactions may be recorded in the same block (condition i), different blocks of the same shard (condition ii), and different shard chains (condition iii). Therefore, the processing order $\prec_\mathbf{P}$ defines a relation for two transactions based on the time \emph{when transactions are processed}. If $TX^1_{\mapsto \epsilon} \prec_\mathbf{P} TX^2_{\mapsto \epsilon}$, we say $TX^1_{\mapsto \epsilon}$ has been processed before $TX^2_{\mapsto \epsilon}$.

\begin{definition}[Execution order]\label{definition-finalization-order}
    Let $\prec_\mathbf{E}$ denote an execution order relation in a sharded system, $TX_{\mapsto \epsilon}$ denote a transaction that calls contract $\epsilon$, $TX_{\mapsto {\epsilon, \epsilon_k}}$ denote a sub-transaction generated to call contract $\epsilon_k$ when executing $\epsilon$, and $\Upsilon(TX_{\mapsto \epsilon})$ be the set of contracts that are called when executing $\epsilon$. Given two transactions $TX^1_{\mapsto \epsilon_1}$ and $TX^2_{\mapsto \epsilon_2}$, we define $TX^1_{\mapsto \epsilon_1} \prec_\mathbf{E} TX^2_{\mapsto \epsilon_2}$ if: for every $\epsilon_k \in \Upsilon(TX^1_{\mapsto \epsilon_1}) \cap \Upsilon(TX^2_{\mapsto \epsilon_2})$, sub-transaction $TX^1_{\mapsto {\epsilon_1, \epsilon_k}}$ is executed before $TX^2_{\mapsto {\epsilon_2, \epsilon_k}}$. 
\end{definition}

The execution order $\prec_\mathbf{E}$ defines a relation only for two transactions involving the same contracts, i.e., transactions exist dependencies. For example, we assume that transaction $TX^1_{\mapsto \epsilon_1}$ calls contracts $\epsilon_1$, $\epsilon_3$, and $\epsilon_4$ respectively with sub-transactions $TX^1_{\mapsto {\epsilon_1, \epsilon_1}}$, $TX^1_{\mapsto {\epsilon_1, \epsilon_3}}$, and $TX^1_{\mapsto {\epsilon_1, \epsilon_4}}$. Similarly, another $TX^2_{\mapsto \epsilon_2}$ calls contracts $\epsilon_2$, $\epsilon_3$, and $\epsilon_4$ respectively with sub-transactions $TX^2_{\mapsto {\epsilon_2, \epsilon_2}}$, $TX^2_{\mapsto {\epsilon_2, \epsilon_3}}$, and $TX^2_{\mapsto {\epsilon_2, \epsilon_4}}$. If $TX^1_{\mapsto {\epsilon_1, \epsilon_3}}$ is executed before $TX^2_{\mapsto {\epsilon_2, \epsilon_3}}$, and $TX^1_{\mapsto {\epsilon_1, \epsilon_4}}$ is executed before $TX^2_{\mapsto {\epsilon_2, \epsilon_4}}$, then we have $TX^1_{\mapsto \epsilon_1} \prec_\mathbf{E} TX^2_{\mapsto \epsilon_2}$. 
% Therefore, $\prec_\mathbf{E}$ indicates \emph{the execution order\footnote{Thus, this work uses the terms execution order and finalization order interchangeably.} of transactions}, which affects the final states of related contracts. 
With $\prec_\mathbf{P}$ and $\prec_\mathbf{E}$, we now give a new fairness definition, called finalization fairness, for a sharded system.

\begin{definition}[Finalization fairness]\label{def-finalization-fariness}
    A sharded system is said to possess finalization fairness property on a contract $\epsilon$, if $\epsilon \in \Upsilon(TX^1_{\mapsto \epsilon_1}) \cap \Upsilon(TX^2_{\mapsto \epsilon_2})$ for any given transactions $TX^1_{\mapsto \epsilon_1}$ and $TX^2_{\mapsto \epsilon_2}$, it satisfies: when $TX^1_{\mapsto \epsilon_1} \prec_\mathbf{P} TX^2_{\mapsto \epsilon_2}$, then $TX^1_{\mapsto \epsilon_1} \prec_\mathbf{E} TX^2_{\mapsto \epsilon_2}$.
\end{definition}

The finalization fairness indicates that if a transaction is processed earlier than another transaction, it should be executed and finalized before the other transaction, i.e., \emph{first-processing first-execution}. Note that not all contracts (e.g., non-order-sensitive contracts) need to be provided finalization fairness, thus a sharded system can be considered to satisfy the finalization fairness as long as it provides such a property for an order-sensitive contract.

\subsection{Goals}\label{sec-goal}
We aim to design a cross-shard consensus protocol to prevent a specific front-running attack from happening across shards (i.e., \emph{cross-shard front-running}) as discussed in \S~\ref{section-attack-details}. Note that the attacking behaviors of such a front-running attack are different from previous front-running attacks. The latter happens within a shard (i.e., \emph{intra-shard front-running}) during the period from when transactions appear in the mempool to when transactions are packed into blocks. However, we emphasize that existing solutions of intra-shard front-running are completely compatible with our protocol and can be easily integrated into our protocol (see \S~\ref{sec-discussion} for more discussions).

Obviously, the finalization fairness is the opposite of the front-running attack, where a victim transaction is processed earlier but executed later than a front-running transaction. We next show how Haechi provides finalization fairness property and thus prevents front-running attacks in a sharded system.

%% file: overview.tex
\begin{figure*}
    \centering
    \includegraphics[width=6.5in]{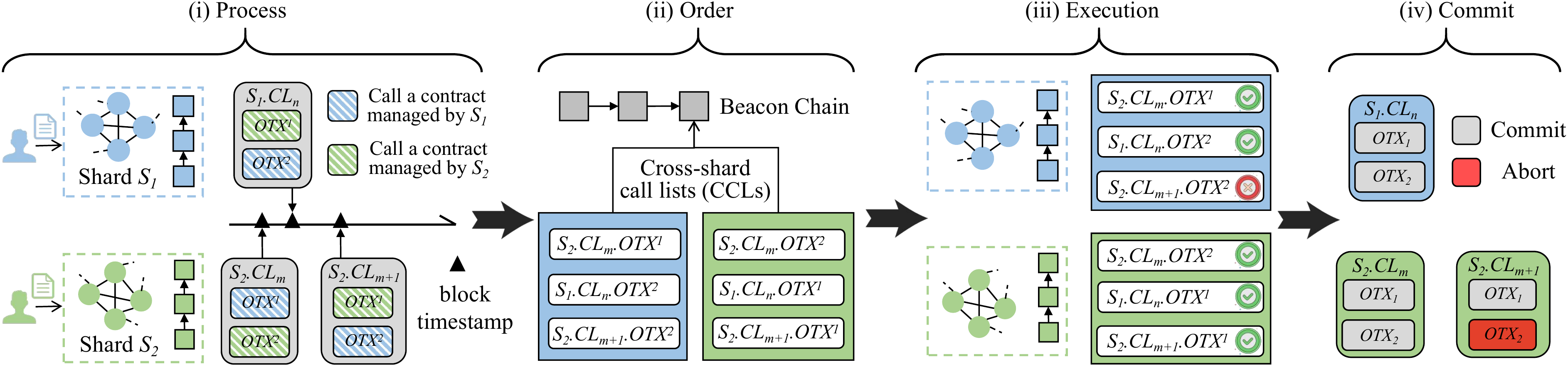}
    \caption{Haechi overview: (i) the sender shards process transactions; (ii) the beacon shard orders transactions; (iii) the contract shards execute contracts; (iv) transactions are committed in senders' and contracts' shards. }
    \label{fig:overview}
% \vspace{-5mm}
\end{figure*}
\section{Haechi Overview}\label{sec-haechi-overview}
In this section, we present an overview and technical challenges of Haechi, followed by a strawman solution. We emphasize Haechi is designed for order-sensitive contracts (OSCs) that are vulnerable to front-running attacks, such as DeFi \cite{sandwich-attack} and ICO \cite{eskandari2019sok}. Therefore, Haechi is implemented as a pluggable cross-shard protocol for the existing sharded systems\footnote{In our implementation, we use a variable to represent different types of contracts, and the system chooses the relevant cross-shard protocol to handle transactions based on the type variable.}. For a clearer illustration, we exclusively consider all involved contracts as OSCs in the rest of this paper. Similarly, if not otherwise specified, all the transactions after that are OSC transactions (OTXs).

% We emphasize Haechi is designed as optional, meaning a contract creator can adopt Haechi to handle the contract that is sensitive to transaction order or adopt classic 2P cross-shard protocol otherwise. This can be easily done in our implementation by adding a type variable to represent different types of contracts, and the system chooses the relevant cross-shard protocol to handle transactions based on the type variable. For distinguishing, we name a contract that is handled by our protocol the \emph{Haechi Smart Contract (HSC)} and use \emph{HSC transaction (HTX)} to represent a transaction that calls a Haechi smart contract.

% \bigbreak
\noindent \textbf{Observation.} Recall that in the front-running attack, an adversary can construct an intra-shard or a cross-shard transaction that is executed earlier than the victim's cross-shard transaction, although the latter is processed first. The main reason is that \emph{no global transaction order is used for intra-shard transactions and cross-shard transactions when they are handled by different shards, respectively}. Motivated by this, Haechi introduces an ordering phase for realizing a globally fair order for OSC transactions.

\subsection{The Process-Order-Execute-Commit Model}\label{sec-poec-model}
Haechi is a process-order-execute-commit (POEC) model that innovatively introduces an ordering phase compared with the 2P cross-shard protocol. We assign the beacon chain to help order transactions in the ordering phase (see more details in \S~\ref{section5-verification}). The data structure used in communication between the beacon chain and shard chains is \emph{CrossLink}. A CrossLink $CL$ is corresponding to one block of a shard chain, consisting of the following components:
\begin{align*}
    CL=<blockTS, L_{tx}, i, h>
\end{align*}
Here, $blockTS$ is the block timestamp of the corresponding block, $L_{tx}$ is a list of OTXs, $i$ is the shard chain ID, and $h$ is the block height of the corresponding block. The beacon chain works in cycles. In each ordering cycle, the beacon chain picks received CrossLinks and orders OTXs of CrossLinks via one instance of the intra-shard consensus.

Fig. \ref{fig:overview} gives an overview of Haechi, which takes the following phases to commit OSC transactions: (i) first, a \emph{processing phase} processes and creates a partial order for transactions, and generates CrossLinks; (ii) then, in an \emph{ordering phase}, the beacon chain decides a total order of transactions in received CrossLinks; (iii) next, in an \emph{execution phase}, the ordered transactions are sent to related shards to execute OSCs; (iv) finally, transactions are committed or aborted by their relevant shards based on the execution results. For distinction, in the following sections, we also name a shard that processes a transaction as the sender shard and a shard that executes the called contract as the contract shard of the transaction.

In a nutshell, Haechi achieves finalization fairness by consistently ordering intra-shard and cross-shard transactions
% \footnote{We emphasize these transactions can be constructed by smart contracts, working as \emph{contract interaction transactions} \cite{ethereum-interaction}. In this case, the sender is a contract and the processing phase will check and modify the state of this contract; while the called contract is still order-sensitive. Since contracts are also a type of account in blockchains \cite{ethereum-account}, such transactions generated by contracts are compatible with our protocol.} 
before they execute the same OSC. As a result, Haechi eliminates the processing-execution difference on those transactions. The question now is how to establish a globally fair order in a sharded system, i.e., transactions that are processed first must be ordered before those that are processed later, regardless of which shards they come from.
% The word \emph{fair} here indicates first processing first execution.

\subsection{Challenges}\label{sec-challenges}
\begin{figure}[t]
    \setlength\abovecaptionskip{0.2\baselineskip}
    \setlength\belowcaptionskip{-1.0\baselineskip}
    \centering
    \includegraphics[width=3.0in]{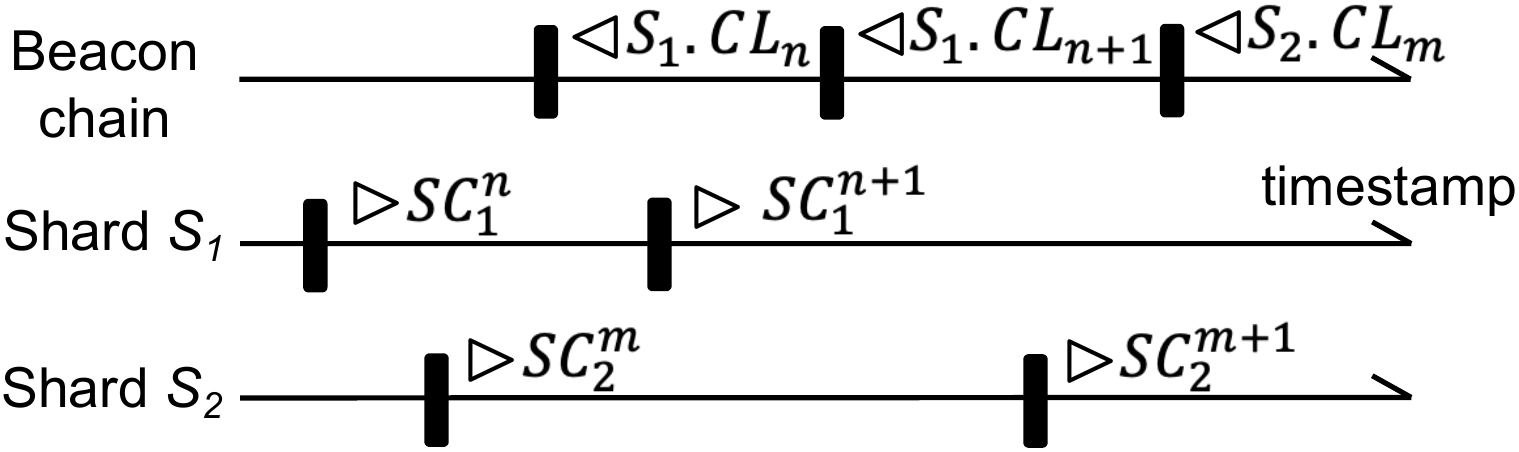}
    \caption{The consensus difference: shards $S_1$ and $S_2$ have distinct consensus speeds, where $\rhd$ indicates the start time of processing a block and $\lhd$ indicates the time when the beacon chain receives the corresponding CrossLink.}
    % , and $|\rhd SC_i^{j+1} - \rhd SC_i^{j}|$ indicates the block interval of shard $S_i$.}
    \label{fig:consensus-difference}
    % \vspace{-5mm}
\end{figure}

In a sharded system, however, establishing a globally fair order is non-trivial. A protocol must ensure that all \emph{in-flight transactions} from different shards can be ordered fairly. Specifically, because of the consensus independence and node heterogeneity of shards (e.g., diverse computer power of nodes, network communication, the number of transaction requests, etc.), different shards have different consensus speeds. Thus, it is impossible for all shards to commit transactions synchronously. Furthermore, due to network delays, the beacon chain may receive transactions from various shards in an order different from their processing time. In some cases, there exist some called in-flight transactions that are being processed by shards or delivered to the network, but the beacon chain has not yet received their corresponding CrossLinks. Such in-flight transactions may be processed earlier than those transactions that the beacon chain has received, and thus should be ordered before to ensure the finalization fairness. For instance, consider a scenario in Fig. \ref{fig:consensus-difference}. Note that the message structure between shards and the beacon chain is the CrossLink, labeled with $S_i.CL_j$, where $i$ is the shard ID, and $j$ is the block height of shard $S_i$.

\noindent\emph{Illustrative Example:} Assume the beacon chain starts to order transactions after receiving $S_1.CL_{n+1}$ from shard $S_1$ (i.e., at $\lhd S_1.CL_{n+1}$ in Fig. \ref{fig:consensus-difference}). Since the beacon chain receives the corresponding CrossLink of $SC_2^{m}$ at the time $\lhd S_2.CL_{m}$ and $\lhd S_2.CL_{m} > \lhd S_1.CL_{n+1}$, the in-flight CrossLink $S_2.CL_{m}$ cannot be ordered in the same ordering cycle as $S_1.CL_{n+1}$. This will lead to the transactions in $SC_1^{n+1}$ being ordered before the transactions in $SC_2^{m}$. However, the transactions in $SC_2^{m}$ are processed earlier than $SC_1^{n+1}$ because $\rhd SC_2^{m} < \rhd SC_1^{n+1}$. As a result, the finalization fairness is violated.

\subsection{Strawman: A Synchronization Ordering Mechanism}\label{sec-strawman}
To address the above challenges, an intuitive solution is to eliminate the in-flight transactions by introducing a synchronization ordering mechanism. Specifically, when a prompt shard commits a block (e.g., assume the block height is $b$) and forwards the relevant CrossLink to the beacon chain, the shard will not propose a new block with height $b+1$ until the beacon chain receives all CrossLinks with the height $b$ from all shards. In other words, the prompt shards need to wait for the sluggish shards before entering the next round of intra-shard consensus. By doing this, all shards synchronize to the same block height regardless of their consensus speeds. As a result, there are no in-flight transactions that will be ignored in each ordering cycle, and all transactions will be ordered fairly based on their processing time no matter which shards they come from.

Although synchronizing blocks among shards can eliminate potential in-flight transactions and ensure a globally fair order, such a synchronization ordering mechanism is low-efficiency and impractical. In particular, in each ordering phase, only one block's transactions are ordered for each shard. The system performance, therefore, will be limited by the slowest shard, which violates the scalability inherence of sharded systems.

%% file: haechidetails.tex
\section{Haechi Details}\label{section: architecture}
In this section, we detail Haechi, where we show how Haechi solves the above challenges. 
% in \cref{section:haechi details-ordering phase}. 
\subsection{Processing Phase}\label{section5-verification} 
The processing phase of Haechi is comparable to the first step of the vote phase in 2P cross-shard protocols where it involves manipulating the senders' accounts and checking their constraints. This is the phase when OTXs are first handled and packed into blocks. 
% Note that this phase should be completed by each shard accountable for managing transactions' senders since it only involves senders' data, e.g., \emph{msg.value} as we discussed in \cref{section-preliminaries-example}. 
Since Haechi decouples the transaction processing, execution, and commitment, we introduce a new block structure motivated by work \cite{ACE} for efficiency reasons.

\noindent\textbf{Haechi block structure.} Blocks in Haechi consist of two types of transaction lists: \emph{pending list} and \emph{committing list}. The pending list records OTXs that have been processed but have not been committed in the current block height. The committing list is used to exclusively pack transactions that can be committed, including non-OSC transactions (i.e., non-contract transactions or contract transactions that don't invoke OSCs) and OTXs that have been executed successfully and are ready for commitment. Such a two-list block structure benefits the efficiency of Haechi.

First, processing and committing OTXs in distinct lists of blocks enables Haechi to achieve \emph{asynchronous processing}, meaning that OTX processing and execution can be completed asynchronously in different relevant shards. Specifically, with the two-list block structure, OTXs can first be processed and recorded in the pending lists, and the sender shards temporarily hold the data changes. After the OTXs are executed by the contract shards, the committing lists of the later blocks are used to record the execution results of relevant contracts and commit previous pending data changes. This enables asynchronous processing for an OTX where the transaction is processed in the sender shard and is executed in the contract shard asynchronously. The asynchronous processing avoids the sender shards from fetching contract states from the contract shards and executing contracts on their own. 

Second, ordering transactions in the pending list before executing them enables Haechi to achieve \emph{non-blocking processing}, meaning that shards can continuously process new OTXs without waiting for the commitments of previous OTXs. The previous 2P cross-shard protocol needs to lock all relevant data to ensure consistent commitments by preventing conflicting transactions from reading/writing the same data in parallel (see appendix \ref{appendix-blocking-verification} for more details). In this case, a later transaction accessing the same data cannot be processed until the former transaction is committed, leading to blocking processing for these conflicting transactions. In contrast, the two-list block structure enables Haechi to form a local order of a shard for conflicting transactions in the pending list, and this local order will help to finally create a global order among all shards. This global order prevents inconsistent commitments even if there are multiple shards processing conflicting transactions in parallel. As a result, a shard can continuously process new OTXs, and pack more than one transaction calling the same OSC into a shard block in a batching manner.

Algorithm \ref{alg-verification} shows the procedure of the processing phase. Once a shard node receives a new block, it goes through all transactions of the blocks (Lines 2-8): (i) the node verifies new OTXs of the pending list by checking if the senders have sufficient balance to pay for transactions (Line 4). Then a valid OTX is added to a transaction list $L_{tx}$, which is the batching implementation for OTX transmission (Line 6); (ii) the node commits updated states of transactions in the committing list (Line 8, details are in \S~\ref{sec-haechi-commitment}). After the new block is handled, the leader of $S_i$ generates a corresponding CrossLink $CL_h$ with the block timestamp $blockTS$, the transaction list $L_{tx}$, and shard id $i$, block height $h$ (Line 10). $CL_h$ is broadcast to other nodes in $S_i$ for collecting signatures (Line 13). Once the leader collects a quorum of signatures (i.e., a quorum certificate), it sends $CL_h$ to the beacon chain (Line 11).

\begin{algorithm}
    \small
    \caption{Haechi processing phase (for nodes $N_k$ of shard $S_i$) in block $SC_i^h$}\label{alg-verification}
    \begin{algorithmic}[1]
        \State \textbf{let} $\{TX\}_{h}$ is the set of transactions in $SC_i^h$, $CL_h=\emptyset$ is the CrossLink of $SC_i$ in block height $h$
        \ForEach {$tx \in \{TX\}_{h} $}, 
        \If {$tx$ is a new OSC transaction}
        \State $isValid\leftarrow Processing(tx)$ \Comment{withdraw coins}
            \If {$isValid$}
            \State $L_{tx} \leftarrow add(L_{tx}, tx)$
            \EndIf
        \ElsIf{$tx$ is ready for commitment}
        \State Commit $tx$ and update the ledger
        \EndIf
        \EndFor
        \If {$N_k$ is the leader}
        \State $CL_h\leftarrow <blockTS, L_{tx}, i, h>$
        \State sends $CL_h$ with a quorum certificate to the beacon chain
        \ElsIf{$N_k$ is not the leader}
        \State signs $CL_h$ received from the leader
        \EndIf
    \end{algorithmic}
\end{algorithm}

\subsection{Ordering Phase}\label{section:haechi details-ordering phase}
The principal goal of the consensus phase is to create a globally fair order for OTXs received from the processing phase. Haechi uses the beacon chain as the entity to accomplish the ordering phase. In previous sharded systems, a beacon chain has been designed as a global coordinator used for global configuration maintenance \cite{Omniledger, Rapidchain, eth2-beaconchain} or cross-shard coordination \cite{ahl, elrond, rivet, gearbox}. Due to the property of global coordination, Haechi extends the beacon chain to assist in ordering OTXs in its blocks. This makes our solution adaptive and easy to be integrated into existing sharded systems. But we emphasize that any shard in the system can be designated to globally order OTXs.

Haechi works as a \emph{leader-based} protocol. A potential concern is whether the beacon chain will become the bottleneck and degrade the performance of a sharded system. To evaluate this, we construct related experiments in \S~\ref{section: evaluation}, showing that our solution only suffers modest performance losses because of the low-cost ordering operation in the beacon chain and the batching optimization. Furthermore, since the relevant shard chains provide data availability for transactions, the transaction data in the beacon chain can be pruned once they are committed. Haechi will thus not introduce momentous storage overhead into the beacon chain. Overall, since the beacon chain in Haechi neither executes transactions nor maintains the state, it is feasible to expand the beacon chain to carry through the ordering phase. Besides, we also discuss another \emph{leaderless} implementation of Haechi. However, without the coordination of a leader, the leaderless Haechi brings more communication overheads, which will be discussed in \S~\ref{sec-discussion}.

\noindent
\textbf{Finalization fairness algorithm}. To solve the in-flight transaction challenges in \S~\ref{sec-challenges}, we propose a finalization fairness algorithm without any synchronization mechanism. To achieve this, Haechi first uses the \emph{block timestamp} as the ordering indicator that is piggybacked on CrossLinks of shards. The block timestamp indicates the time of starting processing transactions in blocks and thus can be used to decide the processing order for transactions. On the other hand, since the block timestamp is publicly attached to each block via the intra-shard consensus, it prevents the attackers from creating arbitrary timestamps to interfere with the ordering phase. Furthermore, the block timestamp enables Haechi to create a \emph{deterministic global order}, and thus delicately prevents the cyclic order problem that is notorious in previous fairness algorithms \cite{byzantine-oligarchy,aequitas}. This is because there are no two nodes of the beacon chain receiving different timestamps regarding the same CrossLink. As a result, all nodes can determine a consistent global order based on block timestamps.

Then, we adopt a so-called \emph{at-least-one ordering rule} to guarantee that all in-flight transactions are ordered fairly. The key is to start ordering transactions \emph{only when the beacon chain receives at least one consecutive CrossLink from each shard}. Since the block timestamp is monotonically incremental for each shard chain, a larger block height indicates a larger block timestamp in a shard. Once the beacon chain receives a new CrossLink with the consecutive block height (e.g., the height is $h+1$ where $h$ is the block height of the last received CrossLink) from a shard, it can determine there are no in-flight transactions that are processed earlier than the transactions in the received CrossLink from the shard. When receiving at least one CrossLink from each shard, the beacon chain has a global view of all in-flight transactions. Therefore, the beacon chain definitely learns which transactions can be ordered in the current ordering cycle.

% By doing this, the beacon chain can catch information of all in-flight CrossLinks and ensure they are ordered fairly. Specifically, nodes in the beacon chain maintain the following data structures: (i) $shardCLMsg$: an $m$-sized two-dimension vector of messages where $shardCLMsgs[i]$ maintains all received-but-unsorted CrossLink messages from shard $S_i$ in the current ordering cycle; (ii) $shardLastTS$: an $m$-sized vector where $shardLastTS[i]$ is the block timestamp of the last CrossLink received from shard $S_i$; (iii) $validTSRange$: an interval $[startTS, endTS]$ for block timestamp, limiting which CrossLinks in shardCLMsg can be ordered in the current ordering cycle.

Algorithm \ref{alg-order} illustrates the procedure of our finalization order fairness algorithm. This algorithm runs in each ordering cycle and will create a new beacon chain block containing an $m$-sized Cross-shard Call Lists $CCLs$. Specifically, $CCL[i]$ corresponds to shard $S_i$, consisting of a set of ordered OTXs whose called OSCs are maintained by $S_i$, also as shown in Fig. \ref{fig:overview}. Note that $CCLs$ is equal to a global order when we merge all OTXs in it. The algorithm consists of three steps:
\begin{asparadesc}
    \item[(a)] CrossLinks preparing (Lines 1-11). Every time when the beacon chain $BC$ receives a CrossLink $S_i.CL$ from $S_i$, it checks if $S_i.CL$ has a consecutive block height with the last CrossLink stored in $shardCLs$ (Line 2). Here $shardCLs$ is an $m$-sized two-dimension vector maintaining all received CrossLinks whose block heights are consecutive. Specifically, for the corresponding $shardCLs[i]$ of $S_i$, there are no unreceived CrossLinks of $S_i$ whose block heights are less than the rear of $shardCLs[i]$. This is used to ensure no in-flight CrossLinks skip the current ordering cycle. If $S_i.CL$ doesn't satisfy it, $BC$ will put $S_i.CL$ into the CrossLink pool $CLPool$ (Line 10). Otherwise, $BC$ adds $S_i.CL$ and its followed CrossLinks from $CLPool$ to $shardCLs[i]$ (Lines 4-8). After that, $BC$ updates the latest received timestamp of $S_i$. The latest received timestamp $shardLastTS[i]$ ensures no in-flight transactions from $S_i$ are processed earlier than currently received transactions from $S_i$ since the block timestamp is incremental with block height (Line 9). 
    % Then, $S_i.CL$ is added into a CrossLink list $shardCLs[i]$, where $shardCLs$ stores all received CrossLinks with consecutive block heights from shards.
    \item[(b)] CrossLinks filtering (Lines 12-13). Once receiving at least one consecutive CrossLink from each shard, i.e., $shardCLs$ stores at least one CrossLink for all shards, $BC$ can determine which transactions can be fairly ordered based on the timestamp of their CrossLinks. Specifically, $BC$ chooses those received CrossLinks whose timestamp is no larger than the minimum of the last CrossLinks among shards, i.e., $\min_{1 \leq i \leq m}\{shardLastTS[i]\}$. This ensures all transactions that will be ordered in the current ordering cycle correspond with a smaller timestamp than any in-flight transactions. We will give proof in Lemma \ref{lemma-finalization-fairness-inflight}. Intuitively, the \emph{minimum among all shards} indicates the smallest timestamp before which $BC$ can guarantee no in-flight transactions.
    \item[(c)] Transactions ordering (Lines 14-17). In this step, $BC$ will order all transactions that are chosen in the step of CrossLinks filtering. The ordering rule follows: (i) transactions in a CrossLink with a smaller block timestamp are ordered first; (ii) transactions in the same CrossLink but with smaller indexes in the CrossLink are ordered first. Then, these ordered transactions are appended to the relevant cross-shard call lists $CCLs$ based on what OSC they invoke. $CCLs$ will be used as the content of a new block of the beacon chain and verified via one instance of intra-shard consensus. Finally, each $CCLs[i]$ will be sent to $S_i$ for the next execution phase.
\end{asparadesc}

\begin{algorithm}
% \label{alg-fairness-order}
    \small
    \caption{The finalization fairness algorithm in Haechi}
    \label{alg-order}
    \begin{algorithmic}[1]
        \ASSUME $m$ shard chains
        \INPUT a set of CrossLinks, an $m$-sized vector $shardLastTS$, an $m$-sized two-dimension vector $shardCLs$
        \OUTPUT an $m$-sized $CCLs$ recorded in a new beacon block
        \Statex {\footnotesize \color{gray}$\blacktriangleright$  Step $1$: CrossLinks preparing}
        % \While {true}
        \State \textbf{upon} receiving a CrossLink $S_i.CL$ from shard $S_i$ \textbf{do}
        % \SCOPE
            % \State \hspace*{5mm} \textbf{if} {$S_i.CL$ has a consecutive block height} \textbf{then}
            \State \hspace*{4mm} \textbf{if} {$S_i.CL.h = shardCLs[i].rear.h+1$} \textbf{then} \\\Comment{{\footnotesize \color{gray}$S_i.CL$ has a consecutive block height of the latest added CrossLink}}
            % \State \hspace*{8mm} add $S_i.CL$ and all CrossLinks from the CrossLink pool with consecutive block heights to $shardCLs[i]$
            \State \hspace*{8mm} $shardCLs[i]$ $\leftarrow add$ ($shardCLs[i]$, $S_i.CL$)
            \\\Comment{{\footnotesize \color{gray} packs all CrossLinks with block heights that are consecutive with that of the last CrossLink in $shardCLs[i]$}}
            \State \hspace*{8mm} \textbf{for all} {$CL \in CLPool$} \textbf{do}
            \State \hspace*{12mm} \textbf{if} {$CL.h = shardCLs[i].rear.h+1$} \textbf{then}
            \State \hspace*{16mm} $shardCLs[i]$ $\leftarrow add$ ($shardCLs[i]$, $CL$)
            \State \hspace*{8mm} $shardLastTS[i]$ $\leftarrow$ $shardCLs[i].rear.blockTS$
            % \State \hspace*{8mm} $shardCLs[i]$ $\leftarrow add$ ($shardCLs[i]$, $S_i.CL$)
            \State \hspace*{4mm} \textbf{elseIf} {$S_i.CL.h > shardCLs[i].rear.h+1$} \textbf{then}  
            \State \hspace*{8mm} add $S_i.CL$ to the CrossLink pool $CLPool$
            % \EndIf
        % \ENDSCOPE
        % \EndWhile
        
        {\footnotesize \color{gray}\Statex $\blacktriangleright$ Step $2$: CrossLinks filtering}
        % \If{receives at least one CrossLink from each shard} 
        \State \hspace*{4mm} \textbf{if} {$|shardCLs[i]|\geq 1, \forall 1 \leq i \leq m$} \textbf{then}
        % \Comment{$BC$ receives at least one CrossLink from each shard}
        \State \hspace*{8mm} Choose all CrossLinks from $shardCLs$
        with a block timestamp $blockTS \leq \min_{1 \leq i \leq m}\{shardLastTS[i]\}$.
        % \EndIf
        
        {\footnotesize \color{gray} \Statex $\blacktriangleright$ Step $3$: Transactions ordering}
        \State \hspace*{4mm} \textbf{for each} {$CL$ chosen in $Step 2$} \textbf{do}
            \State \hspace*{8mm} \textbf{for each} {$OTX \in CL$} \textbf{do}
                \State \hspace*{12mm} \textbf{if} {$OTX$ call an OSC maintained by $S_i$} \textbf{then}
                \State \hspace*{16mm} $CCLs[i]\leftarrow addWithOrder(CCLs[i], OTX)$
                % \Comment{order $OTX$ based on $blockTS$ and the indexes of $OTXs$ in a block}
                % \EndIf
            % \EndFor
        % \EndFor
    \end{algorithmic}
\end{algorithm}
% \vspace{-3mm}

Haechi reduces performance loss caused by the global ordering phase from two aspects. First, our finalization fairness algorithm maintains the transaction throughput by allowing shards to continuously process and commit transactions. As opposed to the strawman solution discussed in \S~\ref{sec-strawman}, shards in Haechi are not blocked and impeded by slow-running shards and can commit those transactions irrelevant to OSCs via intra-shard consensus independently. Haechi, therefore, does not sacrifice the transaction throughput. Second, the at-least-one ordering rule reduces the average confirmation time of OTXs by enabling multiple CrossLinks to be ordered in one ordering cycle. Specifically, those CrossLinks generated by the shards with a fast consensus speed may have small timestamps, and the beacon chain will order them in the same ordering cycle if their timestamps satisfy the CrossLinks filtering condition (Lines 12-13). As a result, with more CrossLinks being ordered within an ordering cycle, the average waiting time for each CrossLink to order and commit will decrease.

% Compared with the solution in \cref{sec-strawman} where only one CrossLink per shard can be ordered, our finalization fairness algorithm supports ordering multiple CrossLinks from the same shard in one ordering cycle, which is significantly efficient when there exist sluggish shards in the sharded system.

\subsection{Execution Phase}\label{sec-haechi-execution}
With deterministic transaction orders received from the beacon chain, a shard can execute the related OSCs with the received cross-shard call list. According to the complexity of an OSC, there are two conditions when executing an OSC:

\noindent\textbf{One-shard execution.} If the invoked OSC is a \emph{simple} contract without contract interactions, the shard can execute it independently via an intra-shard consensus. Specifically, consensus nodes use the input data to execute the OSC. Depending on the execution result, the shard labels a tag for each OTX.
\begin{asparadesc}
    \item[(a)~Lock to Commit.] If the OSC is executed successfully, the shard labels the transaction with a \emph{final-and-commit} tag. Simultaneously, it modifies the local OSC state and locks related data.
    \item[(b)~Request to Abort.] If the shard fails to execute the OSC, the transaction cannot be committed and has to be aborted. In this case, the shard labels the transaction with a \emph{final-but-abort} tag and discards all modifications of the OSC state introduced by the transaction.
\end{asparadesc}

\noindent\textbf{Multi-shard execution.} If the invoked OSC is a \emph{complex} contract that invokes other contracts in other shards during its execution, the shard handles the OTX via a collaborative execution way. During the process of the collaborative execution, each related shard executes the relevant contract it maintains. As the one-shard execution, the shard labels a tag for the transaction according to the execution result of the OSC that it maintains.
\begin{asparadesc}
    \item[(a)~Vote for Commit.] If an OSC in a shard is executed successfully, the OTX is labeled with a \emph{progress-and-commit} tag by the shard. Besides, the shard specifies all other OSCs invoked by its OSC (i.e., contract dependency) so that the related shard can determine when to commit the related transaction and contract states. Meanwhile, the shard records the snapshot of the modifications of the OSC state and locks the related data.
    \item[(b)~Vote for Abort.] If, however, a shard fails to execute its OSC, the OTX is labeled with a \emph{progress-but-abort} tag. Similarly, the shard needs to specify the contract dependency for the commitment decision. Besides, the shard discards all modifications of its OSC state.
\end{asparadesc}

After handling each OTX in the received cross-shard call list, the shard collects the execution results (commit or abort), return value (if they exist), and contract dependency (for multi-shard execution) and then packs them into an execution message with an attestation. The execution message is next forwarded to the sender shard
% the beacon chain (\jianting{should be the sender shard}) 
for the final commitment. 

\subsection{Commitment Phase}\label{sec-haechi-commitment}
In the final commitment phase, each transaction sender shard works as the coordinator to help achieve atomic commitment for the transaction. Based on the tags of received execution messages, there are three actions that nodes of the sender shard could take for an OTX:
\begin{asparaitem}
    \item[(a)] If receiving a message with a \emph{final-and-commit} tag, meaning that only one contract involves in the handling of the OTX and the contract is executed finally successfully, nodes coordinate atomic commitment by sending related shards a commitment message with a verification. Specifically, the shard that records the OTX commits the transaction and modifies the sender's account permanently, while another shard that maintains the invoked contract will be notified to unlock the related data and commit modifications of the contract state.
    \item[(b)] If nodes receive a message with a \emph{progress-and-commit} tag, meaning that there are multiple contracts involved in the handling of the OTX, then it needs to wait for receiving all related messages. Specifically, nodes check the contract dependency in the message to determine if they have received all execution results from related shards. Once nodes receive all messages with the \emph{progress-and-commit} tag, they can commit the transaction. The sender shard modifies the sender's account permanently. Meanwhile, all related shards that involve in modifying contracts are informed about unlocking the related data and committing modifications to the contract state.
    \item[(c)] If nodes receive a message with a \emph{final-but-abort} tag or \emph{progress-but-abort} tag, then the sender shard aborts the transaction and notifies all other related shards to discard the modifications of contracts.
\end{asparaitem}

Finally, all OTXs are recorded in the committing list of shard blocks after the commitment phase. 
% regardless of execution results.

%% file: appendics/apdx-proof-v2.tex
\section{Security Analysis}\label{sec-security-proof-detail}
We next prove the finalization fairness, safety, and liveness of Haechi. 
% under the assumption in \S~\ref{section3-trust-model}.

A sharded system is a multi-chain ledger where each shard maintains a blockchain via an intra-shard BFT consensus protocol. Each shard can therefore be considered a BFT State Machine Replication (BFT-SMR). However, since cross-shard transactions can modify the data of more than one shard simultaneously, it is not adequate to use the original safety and liveness property (cf. intra-shard safety and liveness in \S~\ref{section3-trust-model}) of BFT-SMR to completely describe the security of a sharded system. Therefore, we first define safety and liveness for a sharded system.
\begin{definition}[Sharded System Safety]\label{def-system-safety}
    A sharded system is said to satisfy the safety property if: (i) any two honest nodes of the same shard maintain the same prefix ledger; (ii) any two honest nodes from two different shards have the same commitment sequence and same operation (commit or abort) for all cross-shard transactions involving the two shards.
\end{definition}

\begin{definition}[Sharded System Liveness]\label{def-system-liveness}
    Every transaction received by at least one honest node will be eventually handled by relevant shards and get a response from the sharded system.
\end{definition}

\subsection{Finalization Fairness Analysis}\label{appendix-finalization-fairness-proof}
Recall from the definition in \S~\ref{section3-fairness-definition}, finalization fairness indicates that if two transactions are involved in calling the same contract, then their execution orders (i.e., finalization orders) in all contracts they are jointly involved in are the same as their processing order. In Haechi, transaction execution and commitment follow the global order established in the ordering phase. Therefore, we can prove that Haechi provides finalization fairness by showing that the ordering of transactions in the ordering phase follows the processing order strictly. Note that the beacon chain works in ordering cycles, in each of which the beacon chain picks received CrossLinks selectively and orders OTXs in them.

\begin{lemma}\label{lemma-finalization-fairness-inflight}
    In Haechi, the block timestamp of any in-flight CrossLink is larger than all block timestamps of CrossLinks that are ordered in the current ordering cycling.
\end{lemma}
\begin{myproof}{}
    We prove this lemma by comparing: (i) in-flight CrossLinks with the received CrossLinks from the same shard; (ii) in-flight CrossLinks with the received CrossLinks from different shards.

    \noindent (i) First, in our finalization fairness algorithm (Algorithm \ref{alg-order}), the beacon chain starts ordering in the current ordering cycle once it receives at least one CrossLink with a consecutive block height from each shard. On the one hand, \emph{at least one} ensures that the beacon chain will not miss CrossLinks from any shard in the current ordering cycle. On the other hand, the \emph{consecutive block height} ensures that all CrossLinks corresponding to blocks with smaller block heights have been received in the current ordering cycle or ordered in the previous ordering cycles. Therefore, for every shard, all its in-flight CrossLinks have a higher block timestamp than its CrossLinks received by the beacon chain in the current ordering cycle.

    \noindent (ii) Second, assume $S_i.CL_{last}$ is the last received CrossLink and $\{S_i.CL_{flight}\}$ is a set of all in-flight CrossLinks from $S_i$. In the current ordering cycle, the largest timestamp that is allowed to be ordered is the minimum block timestamp of the last received CrossLinks among all shards. Without loss of generality, we assume $S_1.CL_{last}$ has the minimum block timestamp in the current ordering cycle, and there are $m$ shard chains. Then, we have:
    \begin{equation}
    \label{eq-last-block}
        S_1.CL_{last}.blockTS \leq S_i.CL_{last}.blockTS,
    \end{equation}
    where $1 < i \leq m$. Moreover, according to Proof (i) we can know all in-flight CrossLinks of a shard have a higher block timestamp than the shard's last CrossLink that is received by the beacon chain. Thus, for all shards $S_i$ ($1 \leq i \leq m$) and any CrossLink $cl \in \{S_i.CL_{flight}\}$, we have:
    \begin{equation}
    \label{eq-flight-block}
        S_i.CL_{last}.blockTS < cl.blockTS.
    \end{equation}
    
    Finally, from Equations (\ref{eq-last-block}) and (\ref{eq-flight-block}), we can finally show:
    \begin{equation}
        S_1.CL_{last}.blockTS \leq cl.blockTS,
    \end{equation}
    where $\forall cl \in \{S_i.CL_{flight}\}$ and $1 \leq i \leq m$. In other words, the block timestamp of any in-flight CrossLink is larger than all block timestamps of CrossLinks that are ordered in the current ordering cycling.
\end{myproof}

\begin{lemma}\label{lemma-finalization-fairness-order}
    For any two received transactions $OTX^1$ and $OTX^2$, piggybacked by CrossLinks $CL_1$ and $CL_2$ respectively, if (i) $CL_1.blockTS < CL_2.blockTS$, or (ii) $CL_1 = CL_2$ and $f_{idx}(OTX^1) < f_{idx}(OTX^2)$, then Haechi must order $OTX^1$ before $OTX^2$ in the ordering phase.
\end{lemma}
\begin{myproof}{}
    Note that $f_{idx}(\cdot)$ in Lemma \ref{lemma-finalization-fairness-order} is a function returning the indexes of transactions in a block, same as that of Definition \ref{definition-processing-order}. Moreover, conditions (i) and (ii) in Lemma \ref{lemma-finalization-fairness-order} are actually consistent with the ordering rules we defined in the finalization ordering algorithm (cf. \S~\ref{section:haechi details-ordering phase} step (c)). We now prove Lemma \ref{lemma-finalization-fairness-order} by analyzing all possible byzantine behaviors performed by a malicious leader in the beacon chain.
    
    On the one hand, since the ordering phase is driven by an instance of intra-shard consensus, and $f_{idx}(\cdot)$ and block timestamps are publicly verifiable, a new block created by a malicious leader will not be accepted by other consensus nodes if it contains a transaction order violating the ordering rules, e.g., a transaction with a larger index is ordered before another transaction in the same CrossLink. On the other hand, the malicious leader may deliberately drop some CrossLinks it received, leading to transactions in these dropped CrossLinks are not ordered in the appropriate ordering cycles, thus violating the ordering rules. However, consensus nodes can detect such an omission behavior by checking if the selected CrossLinks in the new block have consecutive block heights for each shard and if the last CrossLinks of all shards are considered in the new block. Overall, any byzantine behaviors violating the ordering rules can be detected and prevented. Since the ordering rules are satisfied all the time, Lemma \ref{lemma-finalization-fairness-order} is correct and the proof is done.
\end{myproof}

With Lemmas \ref{lemma-finalization-fairness-inflight} and \ref{lemma-finalization-fairness-order}, we now prove Theorem \ref{theorem-finalization-fairness}:

\begin{theorem}\label{theorem-finalization-fairness}
    Haechi provides the finalization fairness for a sharded system.
\end{theorem}

\begin{myproof}{} %{ of Theorem \ref{theorem-finalization-fairness}}
    Lemmas \ref{lemma-finalization-fairness-inflight} and \ref{lemma-finalization-fairness-order} guarantee all in-flight and received CrossLinks must be ordered based on the ordering rules. Since the block timestamp and transaction indexes in the ordering rules indicate the processing order, the beacon chain will keep this order for the following execution and commit phases. Therefore, Haechi ensures that the processing order is consistent with the execution order, i.e., ensuring the finalization fairness.
\end{myproof}

\subsection{Safety Analysis}
We show that Haechi guarantees the sharded system safety property by respectively showing Haechi satisfies the two conditions in Definition \ref{def-system-safety}.
\begin{lemma}\label{lemma-intra-shard-consistency}
    For every two honest nodes $N_i^1$ and $N_i^2$ with local ledgers $\mathfrak{L_i}^1$ and $\mathfrak{L_i}^2$ in any shard $S_i$, Haechi guarantees: for all block height $h$ ($0\leq h \leq min\{f_{len}(\mathfrak{L_i}^1), f_{len}(\mathfrak{L_i}^2)\}$), $\mathfrak{L_i}^1[h] = \mathfrak{L_i}^2[h]$, where $f_{len}(\cdot)$ computes the current block height of a chain.
\end{lemma}

\begin{myproof}{}
    Since each shard runs a BFT consensus for maintaining its own ledger and each shard is controlled by the honest nodes (see the trust assumptions in \S~\ref{section3-trust-model}), the shard satisfies the intra-shard safety according to the safety properties of BFT-based protocol. Furthermore, Haechi does not violate the security of the intra-shard consensus. Therefore, $N_i^1$ and $N_i^2$, which participate in the same shard, have a consistent ledger in Haechi. 
\end{myproof}

\begin{lemma}\label{lemma-inter-shard-consistency}
    If an OTX involves the modification of several data managed by multiple shards, then all the relevant shards either commit their local modification or abort it.
\end{lemma}
\begin{myproof}{}
    Recall from \S~\ref{section: architecture} that the transaction sender's shard in Haechi is responsible for coordinating the cross-shard consensus of an OTX. Once nodes of the sender's shard receive messages from all shards relevant to the transaction, the sender's shard runs a BFT-based intra-shard consensus to make a decision based on the received messages. Since the sender's shard is controlled by the honest nodes, such a consensus can be completed securely and generate one commitment decision for the transaction. Since there is no equivocation, all relevant shards will receive a consistent commitment decision on whether to commit or abort their local modification for the transaction. Moreover, the finalization fairness provided by Haechi (cf. Theorem \ref{theorem-finalization-fairness}) ensures the transaction is eventually committed or aborted.
\end{myproof}

With Lemmas \ref{lemma-intra-shard-consistency} and \ref{lemma-inter-shard-consistency}, we now prove Theorem \ref{theorem-system-safety}:

\begin{theorem}\label{theorem-system-safety}
    Haechi guarantees the sharded system safety.
\end{theorem}

\begin{myproof}{}
    If the two nodes belong to the same shard, by Lemma \ref{lemma-intra-shard-consistency}, they can keep a consistent ledger via the intra-shard consensus protocol. In other words, these two nodes have the same commitment sequence and operation (i.e., commit or abort) for all transactions recorded in their shard chain. On the other hand, if the two nodes belong to different shards, by Lemma \ref{lemma-inter-shard-consistency}, they can take consistent action to every OSC transaction via Haechi, i.e., either all commit or all abort transactions. Furthermore, since Haechi ensures finalization fairness where the processing order is the same as the commitment order, these two nodes will have the same commitment sequence for all OSC transactions they involve. Together, these observations imply the desired result, i.e., Haechi satisfies the sharded system safety property. 
\end{myproof}

\subsection{Liveness  Analysis}
We now give brief proof to show that Haechi satisfies the liveness property of a sharded system.

\begin{theorem}\label{theorem-system-liveness}
    Haechi guarantees the sharded system's liveness.
\end{theorem}
\begin{myproof} {}
    Based on the type of transaction, there are two conditions for handling a transaction $TX$:

    \noindent (i) $TX$ is a non-OTX. In this case, $TX$ will be handled via the original consensus protocol of the sharded system. Specifically, if $TX$ is an intra-shard transaction of a shard, then the shard is responsible for handling it via the intra-shard consensus protocol. If $TX$ is a cross-shard transaction, then all relevant shards collaboratively handle it via the 2P cross-shard consensus protocol. But in either case, the user can receive a response from the sharded system, which is ensured by the underlying protocol of the sharded system.

    \noindent (ii) $TX$ is an OTX. As messages between two honest parties cannot be dropped, the beacon chain eventually receives a CrossLink containing $TX$. Besides, the finalization fairness provided by Haechi guarantees that there is no CrossLink being dropped (otherwise, it violates the finalization fairness since transactions in the dropped CrossLinks cannot be ordered fairly). Since an honest shard is responsive according to the intra-shard liveness provided by the intra-shard BFT protocol. Haechi ensures that any OTX is handled and can get a response eventually.
    
    According to the above analysis, we can conclude that Haechi satisfies the sharded system liveness property.
\end{myproof}

%% file: evaluation.tex
\begin{figure*}[htbp]
	\centering
    \begin{minipage}[t]{0.19\textwidth}
        \centering
        \includegraphics[width=1.4in]{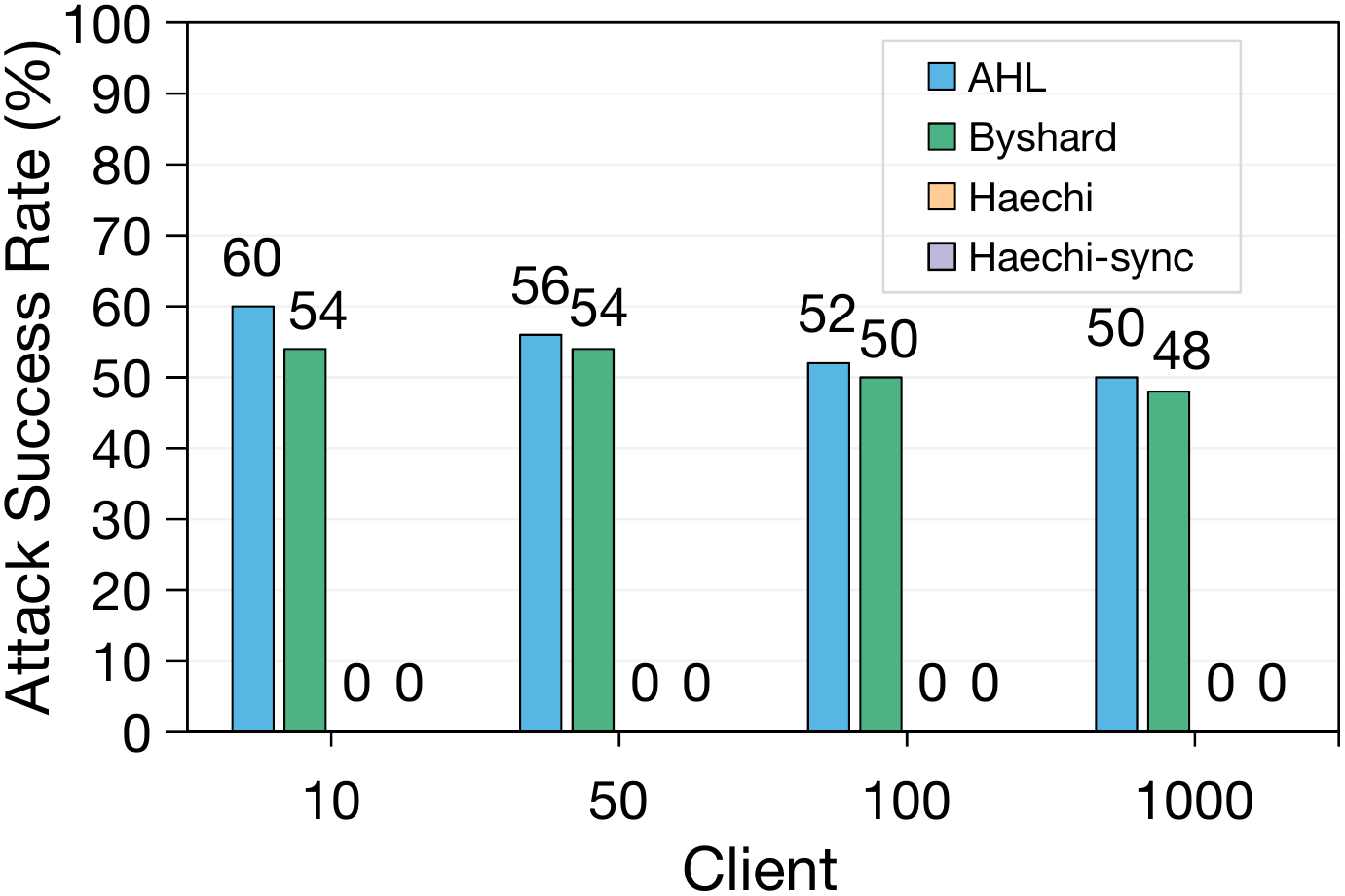}
        \caption{Attack success rate under varying workloads.}
        \label{fig-client-attack}
    \end{minipage}
    \hfill
    \begin{minipage}[t]{0.19\textwidth}
		\centering
		\includegraphics[width=1.4in]{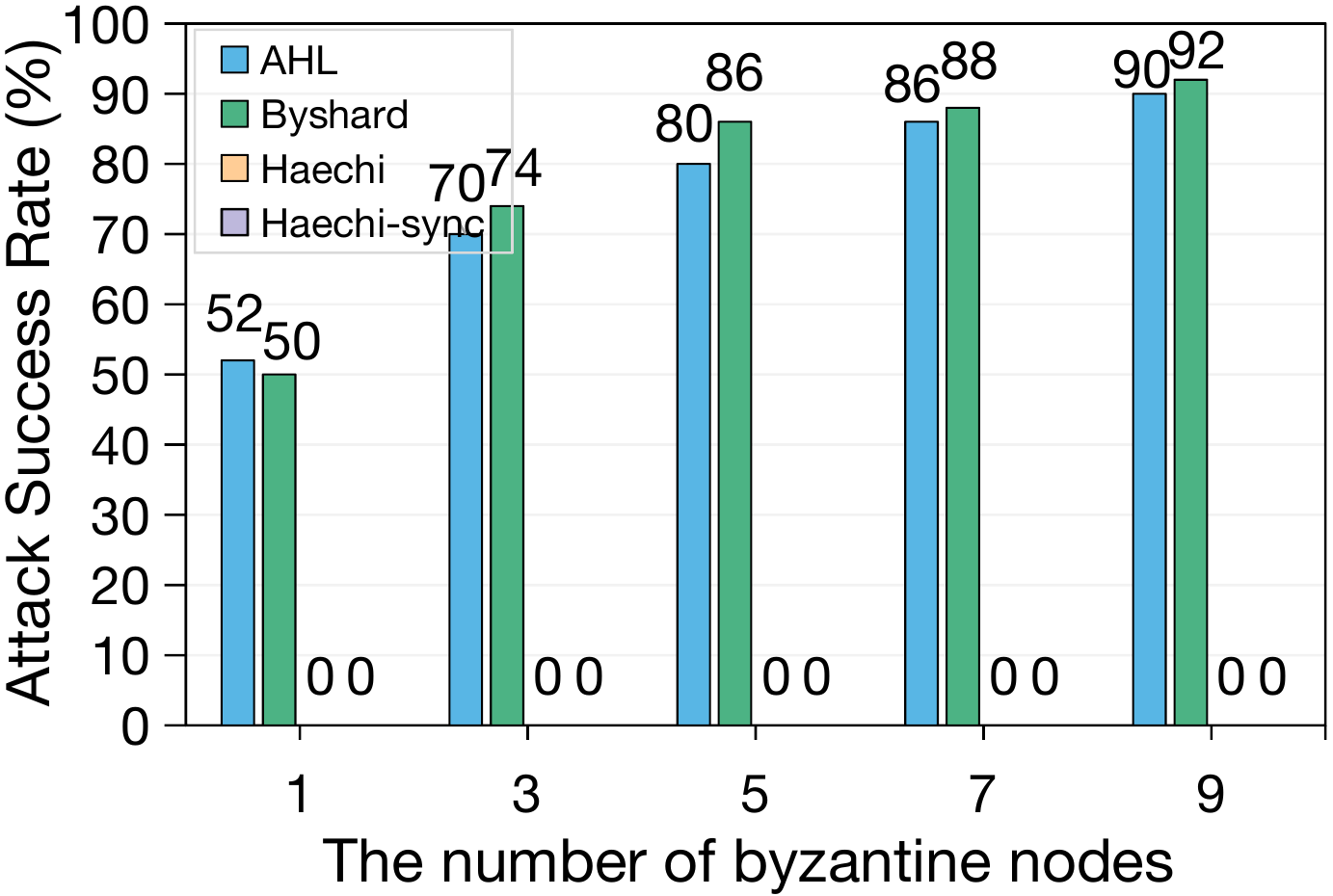}
		\caption{Attack success rate under varying byzantine ratios.}
		\label{fig-byzantine-attack}
	\end{minipage}
    \hfill
	\begin{minipage}[t]{0.19\textwidth}
		\centering
		\includegraphics[width=1.4in]{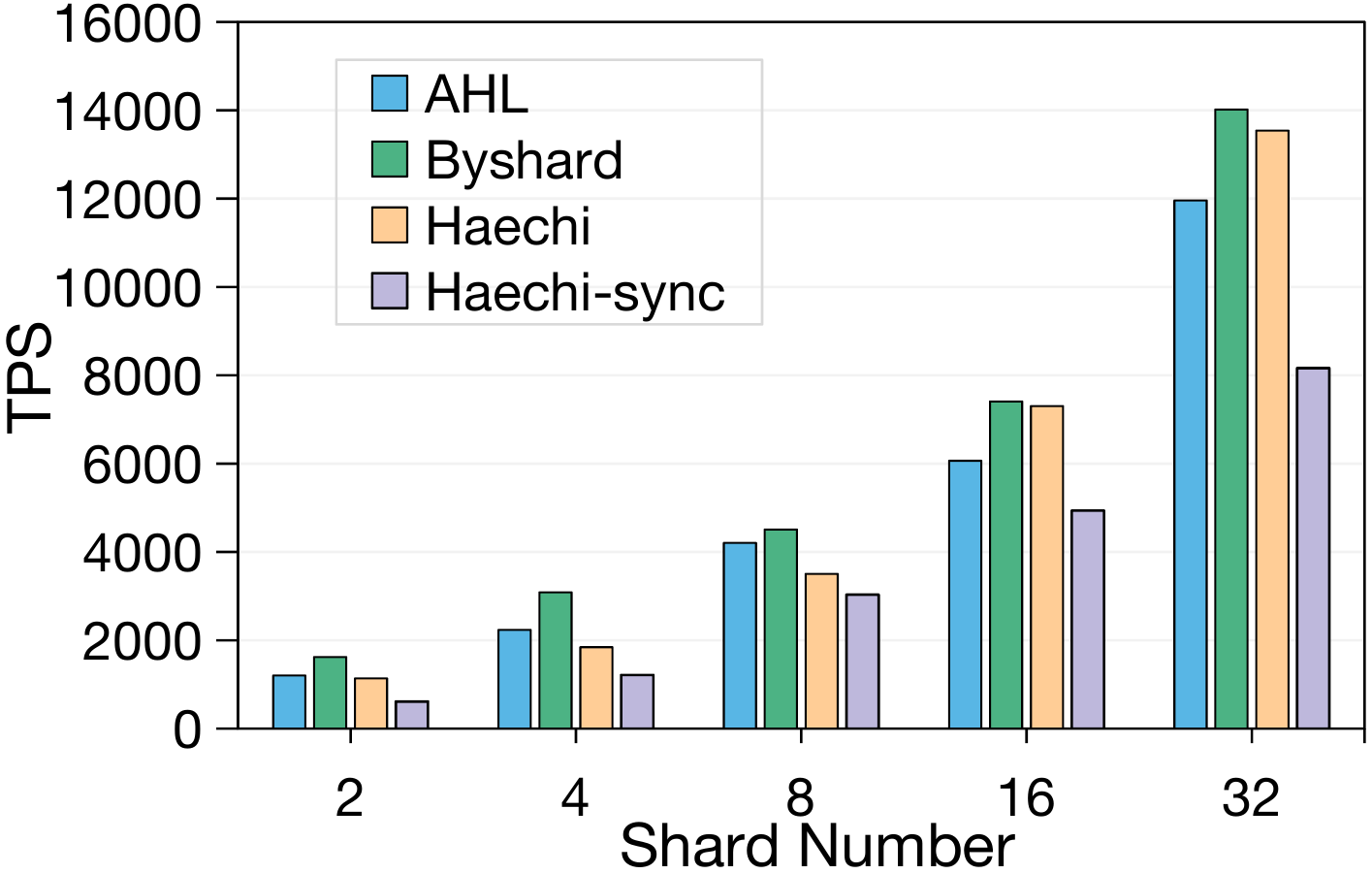}
		\caption{Throughput under various numbers of shards.}
		\label{fig-tps-shard}
	\end{minipage}
	\hfill
	\begin{minipage}[t]{0.19\textwidth}
		\centering
		\includegraphics[width=1.4in]{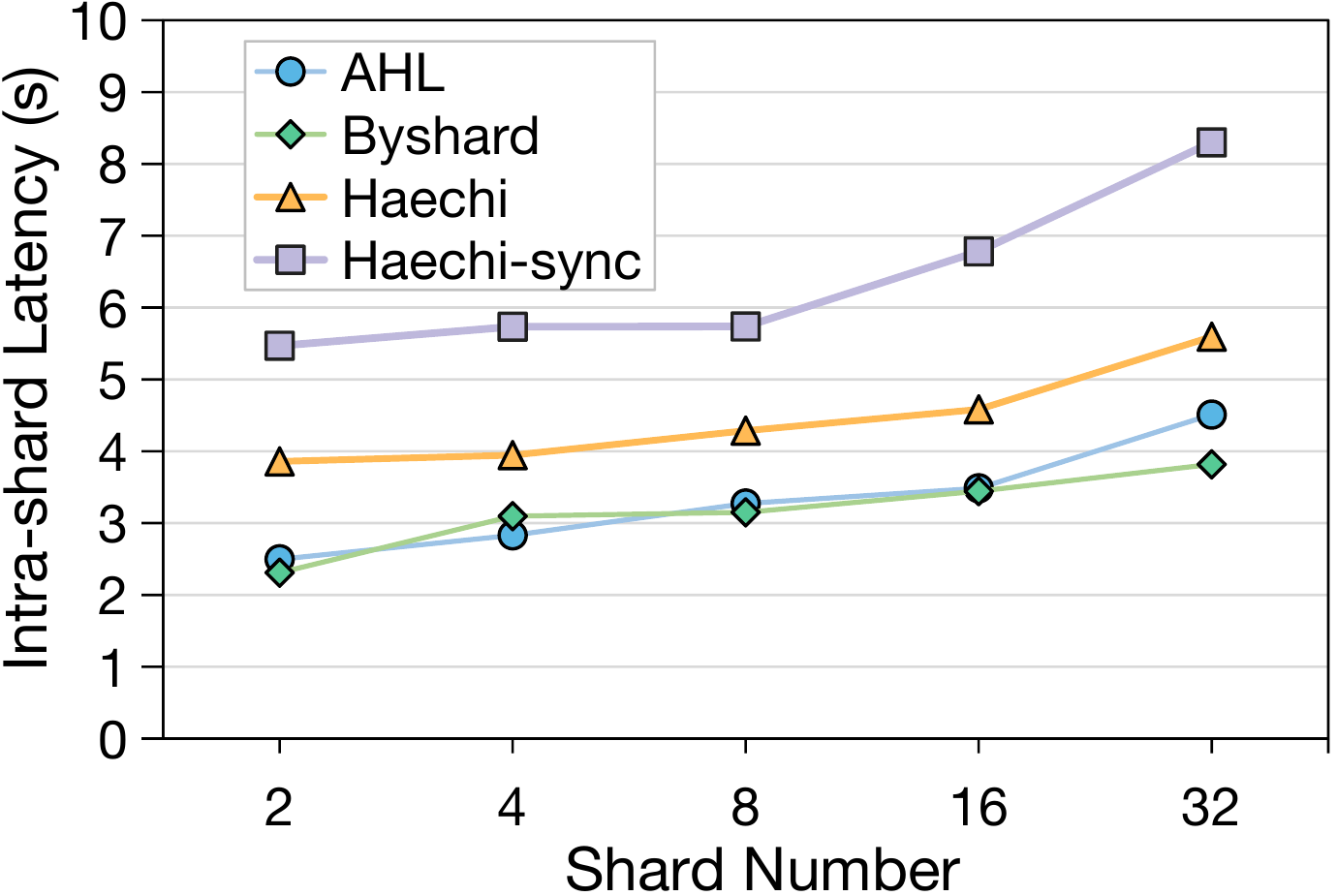}
		\caption{Confirmation latency of intra-shard transactions.}
		\label{fig-intra-latency-shard}
	\end{minipage}
	\hfill
	\begin{minipage}[t]{0.19\textwidth}
		\centering
		\includegraphics[width=1.4in]{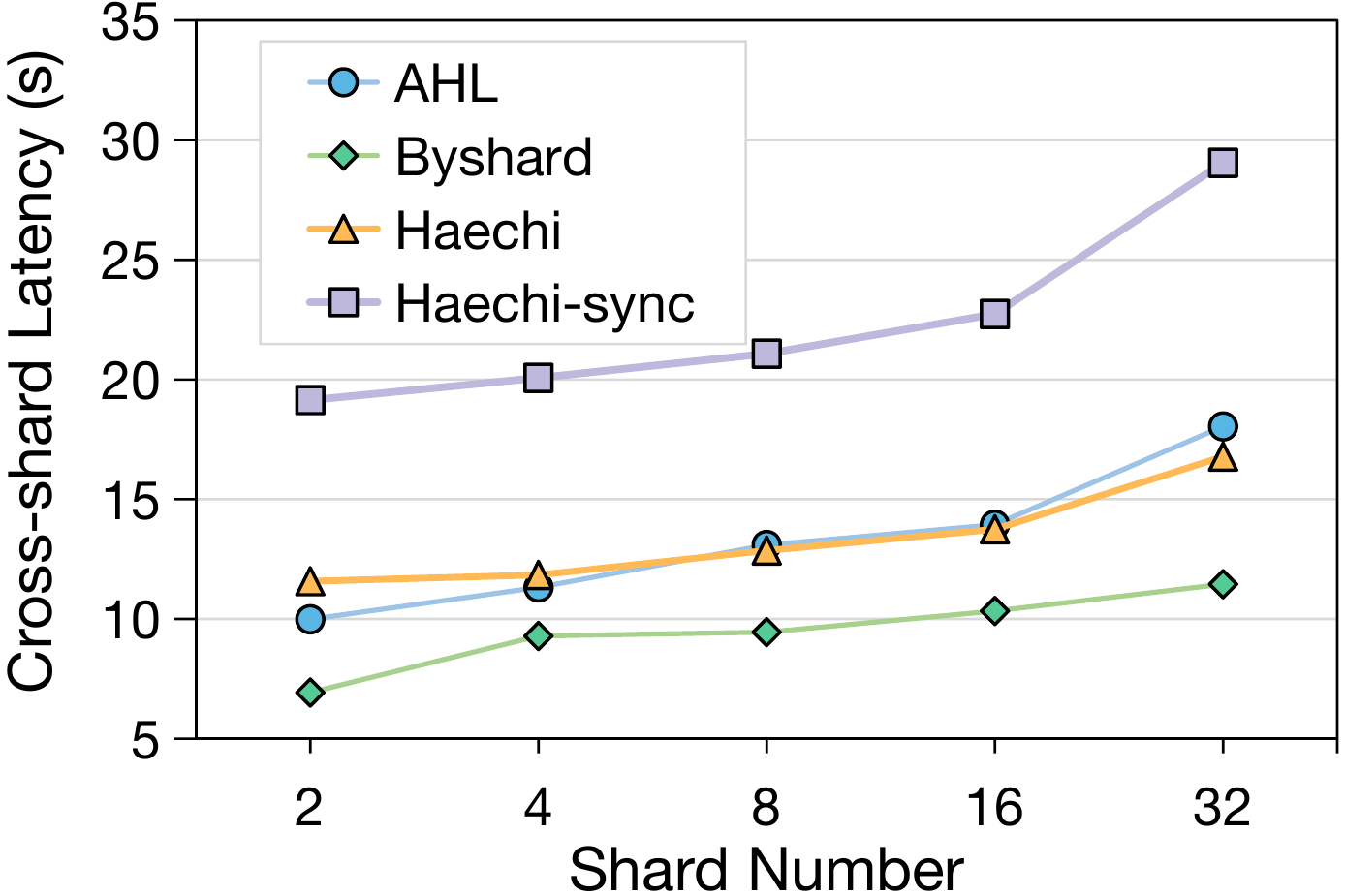}
		\caption{Confirmation latency of cross-shard transactions.}
		\label{fig-cross-latency-shard}
	\end{minipage}
% \vspace{-5mm}
\end{figure*}
\section{Evaluation}\label{section: evaluation}
In this section, we will first present implementation details and experimental settings. Then, we evaluate and compare Haechi with other cross-shard protocols in terms of the ability to prevent front-running attacks, practicality, and performance. 

\noindent \textbf{Implementation.} We implement Haechi based on Tendermint \cite{tendermint}. Tendermint provides modular designs for building blockchains, which allows us to smoothly realize our fair ordering algorithm and cross-shard communication without compromising the security of its consensus protocol. In our implementation, each shard runs the Tendermint BFT protocol \cite{tendermint-bft} to reach a consensus on new blocks, and each node of the same shard maintains the same blockchain ledger that stores transaction data in LevelDB. For cross-shard communications, we assign a leader node of each shard as the cross-shard portal, responsible for sending/receiving requests from other shards to avoid redundancy message delivery. New blocks in a shard chain are created by the shard proposer randomly, while the blocks generation in the beacon chain incorporates the finalization fairness algorithm (Algorithm \ref{alg-order}) to achieve fair ordering for OSC transactions across shards. We implement multiple kinds of Clients to securely send transactions via the interface provided by Tendermint, where all transactions will be verified before processing.

% \bigbreak
\noindent \textbf{Baselines for comparison.} For comparison, we implement three cross-shard protocols based on Tendermint: Haechi-sync, AHL~\cite{ahl}, and Byshard~\cite{byshard}. Haechi-sync is the strawman solution we discussed in \S~\ref{sec-strawman}, where shards are blocked and cannot process new transactions until the beacon chain finishes the ordering phase of the last round (i.e., all shards synchronize to the same block height).
AHL and Byshard are two state-of-the-art 2P cross-shard protocols vulnerable to front-running attacks. More importantly, they support smart contract-type transactions that are considered in our scenario. Both of them delegate a shard as the cross-shard coordinator. The main difference between them is that AHL uses a dedicated \emph{reference shard} to coordinate the processing of all cross-shard transactions, while Byshard asks the sender shard to be the cross-shard coordinator. 
The total implementation involves roughly 4.7K lines of code in Golang. In this section, we will evaluate the presented front-running attack and the performance of these cross-shard protocols.
% compare Haechi with these three cross-shard protocols to show the performance.
% \footnote{https://github.com/PUFreedomLab/haechi/tree/tendermint/haechi.}.

% \bigbreak
\noindent \textbf{Setup.} We run our evaluation in a geo-distributed environment spread across 10 AWS regions in the world (4 in America, 3 in Asia, and 3 in Europe), which includes ten c5a.8xlarge EC2 instances and each one has 32 CPUs, 64 GB RAM, and a 10 Gbps network connection. The sender address and receiver's address of a transaction is generated randomly to guarantee each shard has a balancing workload. The number of nodes in a shard (including the beacon shard) is set to 30, and shards have been running for over 3 minutes for each metrics testing. The mempool of each node can store up to 50,000 transactions, with a maximum byte size of 1 GB. We deploy multiple clients for sending transactions.

% \bigbreak
\noindent \textbf{Metrics.} We use the following metrics for evaluation: 
(i) attack success rate: the effectiveness of the cross-shard front-running attack;
(ii) transaction throughput% \footnote{Note that the throughput computation in our evaluation includes those sub-transactions introduced by cross-shard transactions. But since the number of the sub-transactions depends on the ratio of cross-shard transactions, transactions complexity, etc., we omit the discussions for simplicity.}
: the number of transactions a system can execute per second, i.e., TPS in short; (iii) intra-shard latency: the confirmation latency of intra-shard transactions, starting from the time when the client sends an intra-shard transaction to the time when the transaction is committed by a shard; (iv) cross-shard latency: the confirmation latency of cross-shard transactions, starting from the time when the client sends a cross-shard transaction to the time when the transaction is committed by all relevant shards; (v) CCLs interval: the generation interval of cross-shard call lists.

\subsection{Evaluation of Front-running Attack}\label{section-evl-attack}
We first evaluate the effectiveness of our presented cross-shard front-running attack. To achieve this, we implement the front-running attacking behaviors as illustrated in Fig. \ref{fig:attack-details} on all evaluated protocols. Specifically, we implement byzantine nodes that create front-running intra-shard transactions once they observe victims' cross-shard transactions are processed during intra-shard consensus. We then trace the execution order between the front-running transactions and the victims' transactions from log files. We evaluate the effectiveness of the front-running attack with the attack success rate, which represents a ratio of the front-running transactions that are executed before the victims' transactions. 
% In particular, the attack success rate reflects the ability of a cross-shard protocol to prevent cross-shard front-running attacks.

Fig. \ref{fig-client-attack} and \ref{fig-byzantine-attack} show the attack success rate of four compared protocols under different settings, where we create 50 victims' transactions and repeat 5 times for each experimental setting. In Fig. \ref{fig-client-attack}, we evaluate the impact of transaction workloads on the attack success rate. We initiate varying numbers of clients to send transactions to the contract shard. For 2P cross-shard protocol, Fig. \ref{fig-client-attack} demonstrates that with higher workloads on the contract shard, the attack success rate decreases. This is because once a contract shard suffers from a heavy workload, front-running transactions cannot be processed immediately and may thus be processed with victims' cross-shard transactions in the same instance of intra-shard consensus. Fig. \ref{fig-byzantine-attack} shows the impact of the ratio of byzantine nodes on the attack success rate, where we initiate 1, 3, 5, 7, and 9 byzantine nodes (out of 30 nodes) in each shard. The experiments show that if more front-running attackers exist in a shard, the possibility of being front-run will be higher in 2P cross-shard protocols. In contrast, both Haechi and Haechi-sync can prevent such attacks effectively and we did not observe any successful front-running attacks under a realistic test environment.

\subsection{Scalability}\label{section-scalability}
We then evaluate the scalability of different cross-shard protocols, where we compute transaction throughput, intra-shard and cross-shard latency under the number of shards $m=2,4,8,16,32$. To measure TPS, we count the number of valid transactions that are packed and handled in a block via the log files generated by the ledger. To measure confirmation latency, we trace some transactions and compute the delay from when it was sent to when it was committed. 

As shown in Fig. \ref{fig-tps-shard}, \ref{fig-intra-latency-shard}, \ref{fig-cross-latency-shard}, Haechi scales well as the other two 2P cross-shard protocols. The TPS of Haechi increases nearly linearly while the intra-shard and cross-shard confirmation latency increase slowly with the increasing number of shards. Haechi can achieve 13,000+ TPS, ~6s intra-shard latency, and ~17s cross-shard latency with 32 shards. In all cross-shard protocols, the performance of Byshard is the best regardless of the number of shards since it amortizes the overhead of cross-shard communications to every shard. Furthermore, AHL performs worse than Haechi with increasing shard numbers, as it requires one reference shard to receive and coordinate all cross-shard transactions, which will lead to a bottleneck of the reference shard. Haechi-sync also does not scale well because of its synchronization mechanism. In contrast, the ordering phase performed by Haechi does not introduce significant overhead because of the finalization fairness algorithm and low-cost ordering operations.
% In contrast, even though Haechi also requests a beacon chain for transaction ordering, the senders' shards verify these cross-shard transactions first and transmit them to the beacon chain by batching.

\begin{figure*}[htbp]
	\centering
    \begin{minipage}[t]{0.19\textwidth}
        \centering
        \includegraphics[width=1.4in]{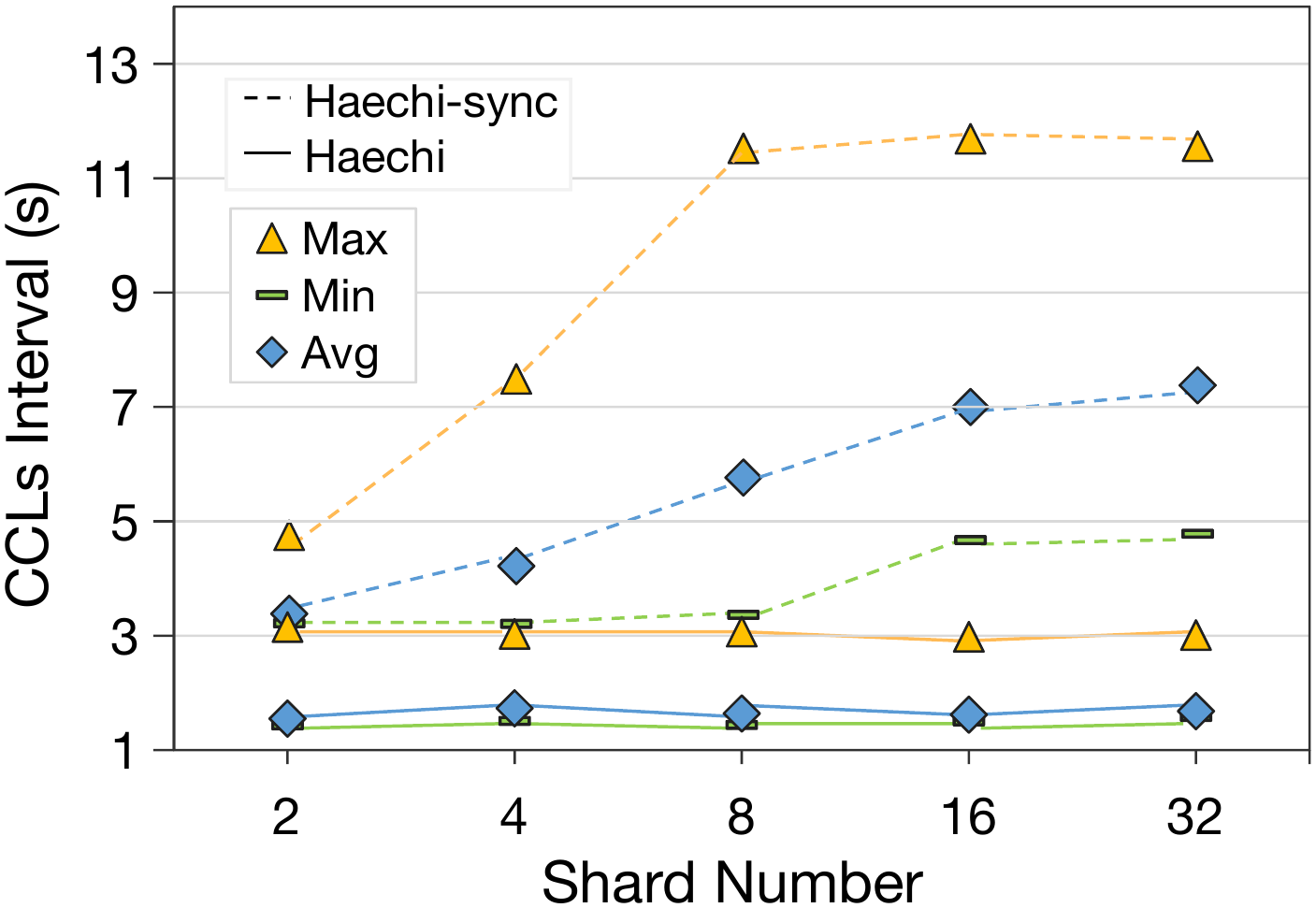}
        \caption{The generation interval of CCLs.}
        \label{fig-shard-ccls}
    \end{minipage}
	\hfill
    \begin{minipage}[t]{0.19\textwidth}
		\centering
		\includegraphics[width=1.4in]{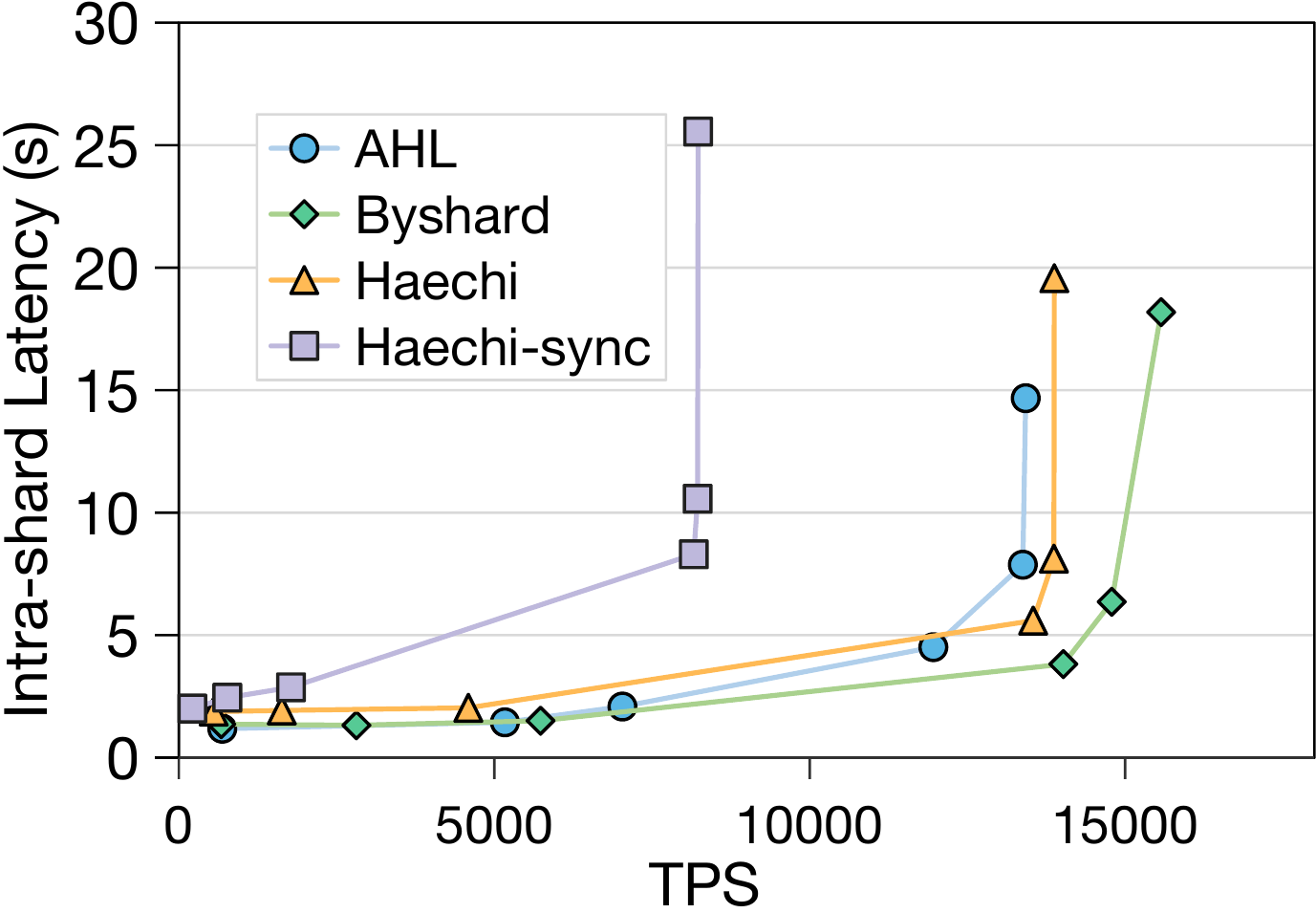}
		\caption{TPS vs. intra-shard latency.}
		\label{fig-tps-intra-shard}
	\end{minipage}
	\hfill
    \begin{minipage}[t]{0.19\textwidth}
		\centering
		\includegraphics[width=1.4in]{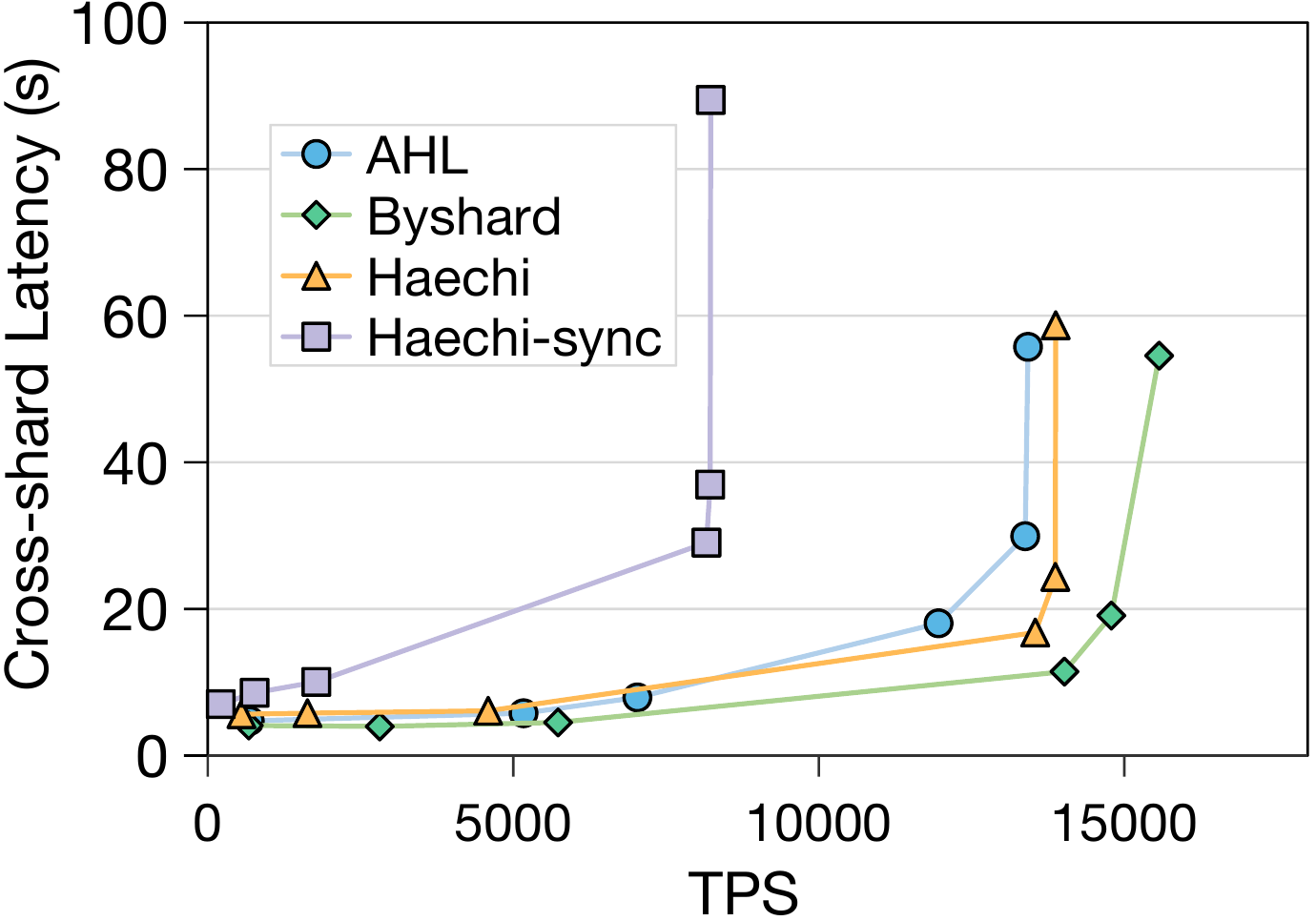}
		\caption{TPS vs. cross-shard latency.}
		\label{fig-tps-cross-shard}
	\end{minipage}
	\hfill
	\begin{minipage}[t]{0.19\textwidth}
		\centering
		\includegraphics[width=1.4in]{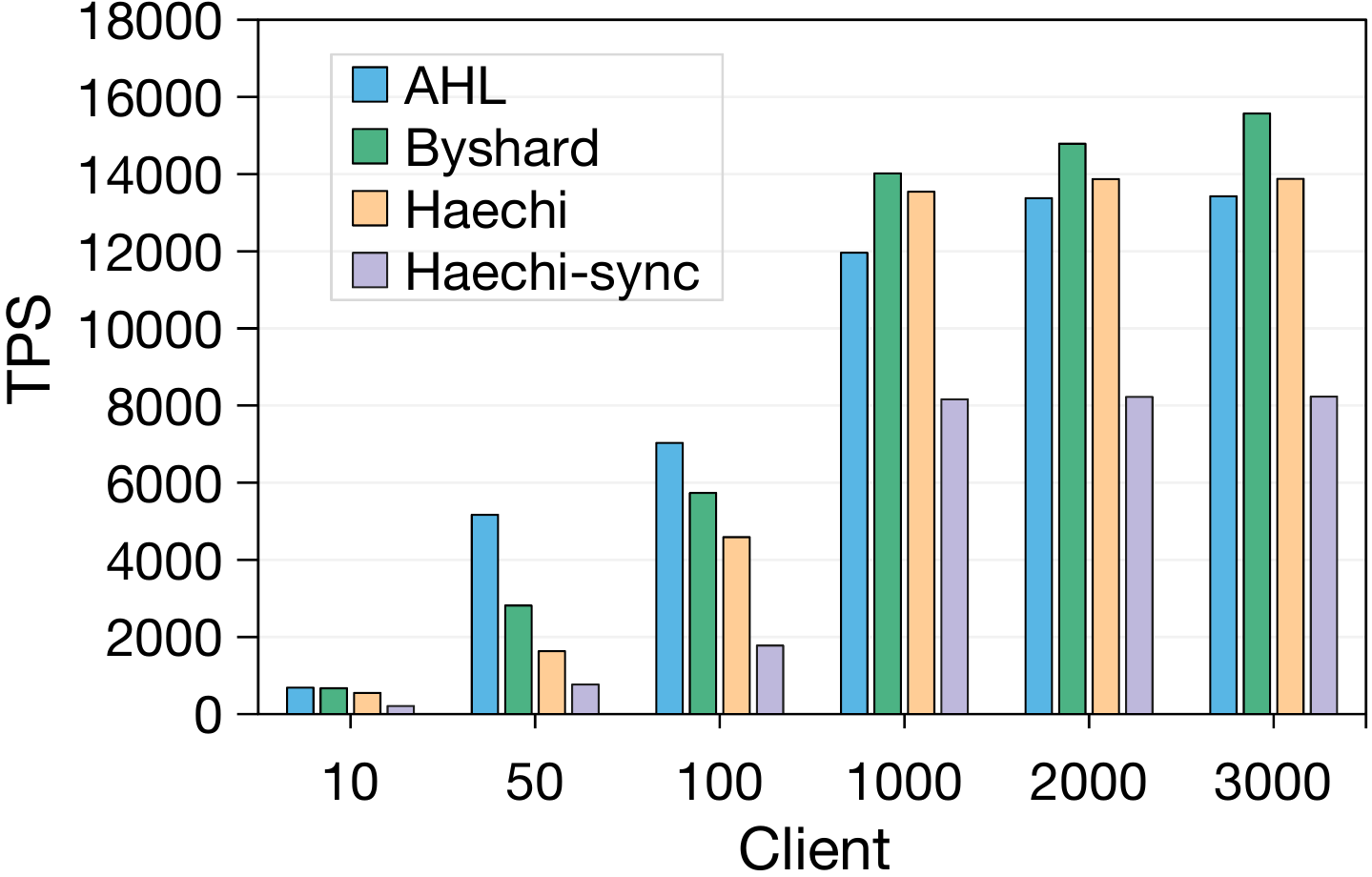}
		\caption{TPS under varying numbers of clients.}
		\label{fig-tps-client}
	\end{minipage}
	\hfill
	\begin{minipage}[t]{0.19\textwidth}
		\centering
		\includegraphics[width=1.4in]{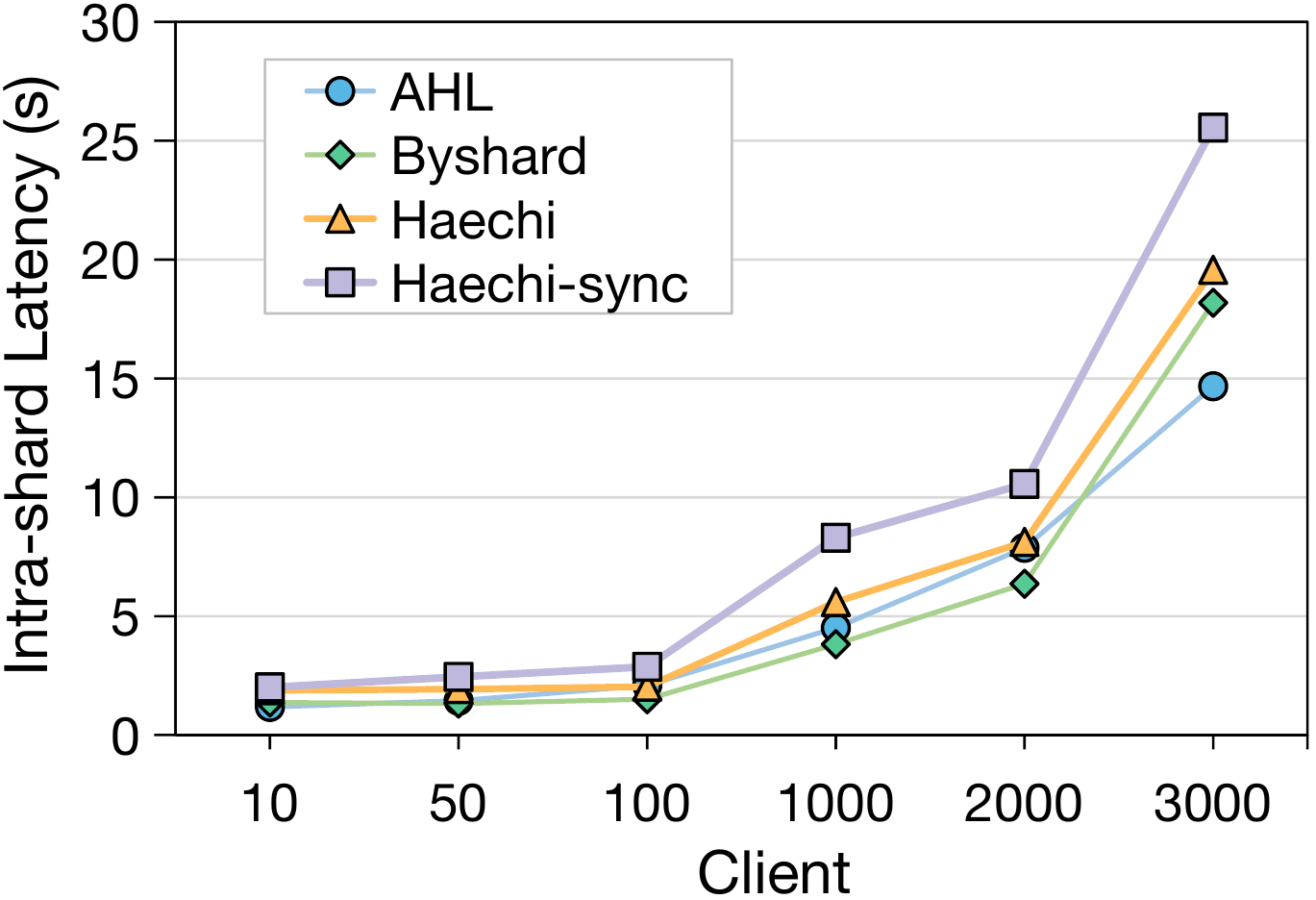}
		\caption{Intra-shard latency under varying clients.}
		\label{fig-intra-latency-client}
	\end{minipage}
	% \hfill
 % \vspace{-4mm}
\end{figure*}

\begin{figure*}[htbp]
	\centering
        \begin{minipage}[t]{0.23\textwidth}
		\centering
		\includegraphics[width=1.7in]{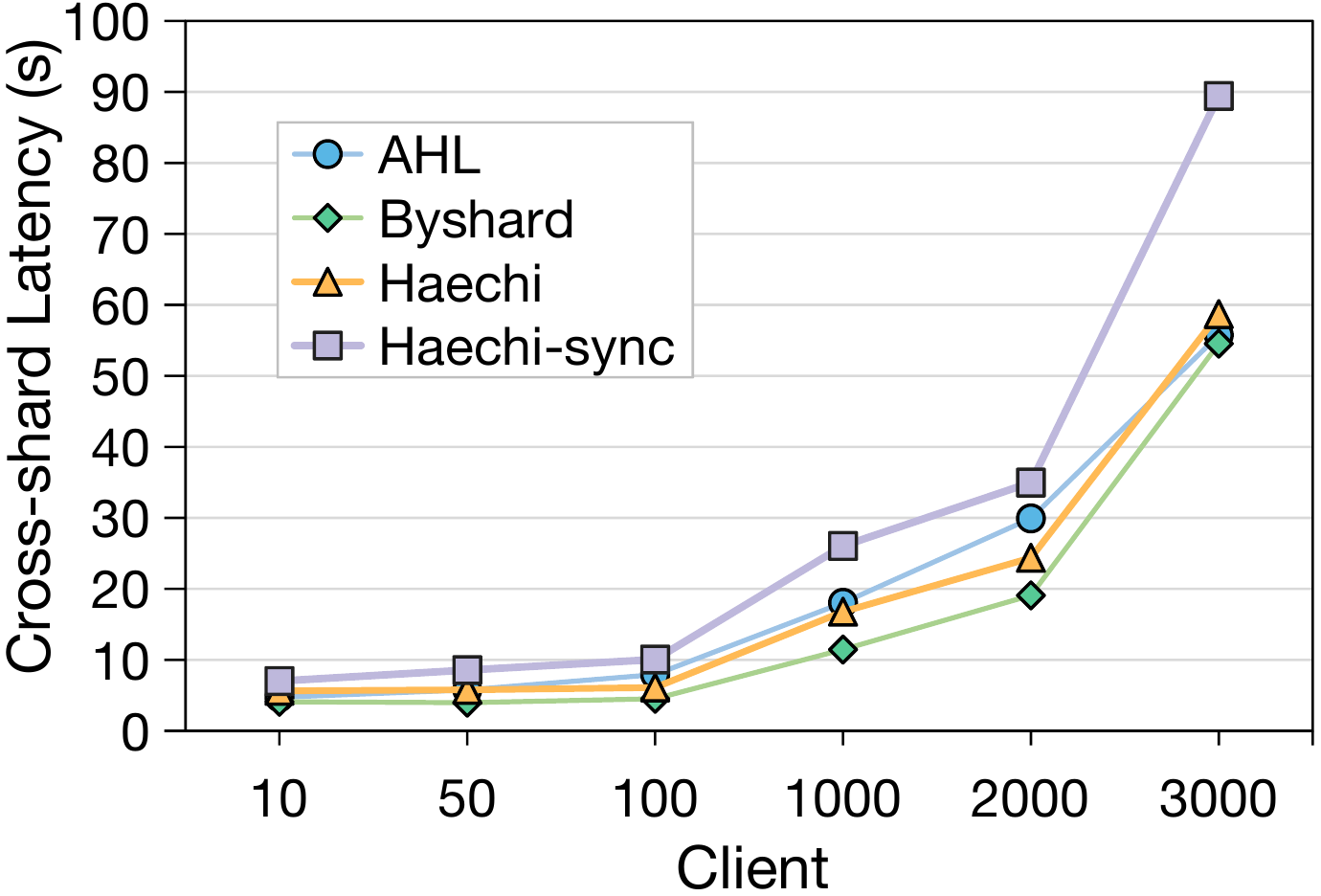}
		\caption{Cross-shard latency under varying clients.}
		\label{fig-cross-latency-client}
	\end{minipage}
        \hfill
	\begin{minipage}[t]{0.23\textwidth}
		\centering
		\includegraphics[width=1.7in]{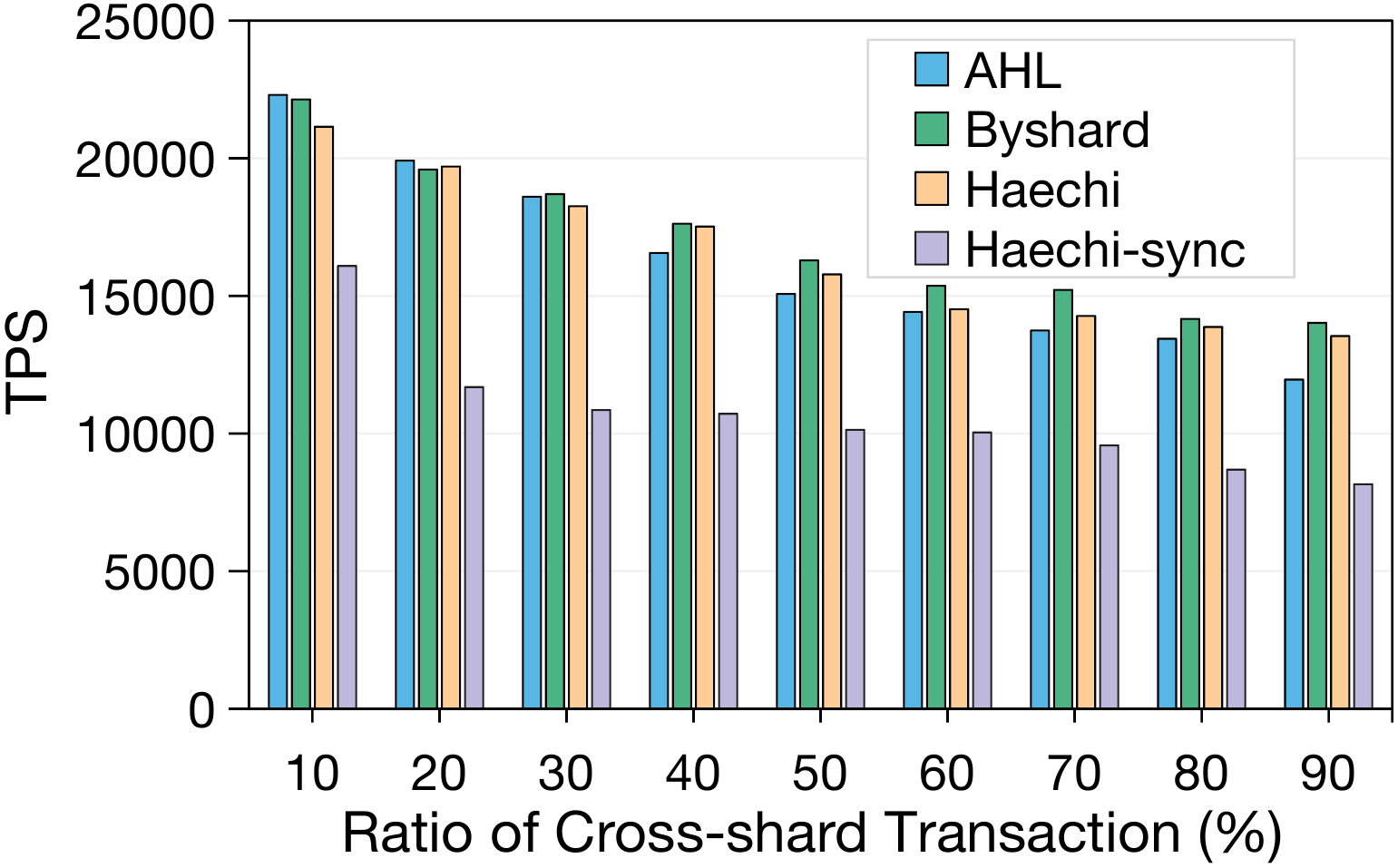}
		\caption{TPS under varying cross-shard ratios.}
		\label{fig-tps-ratio}
	\end{minipage}
	\hfill
	\begin{minipage}[t]{0.23\textwidth}
		\centering
		\includegraphics[width=1.7in]{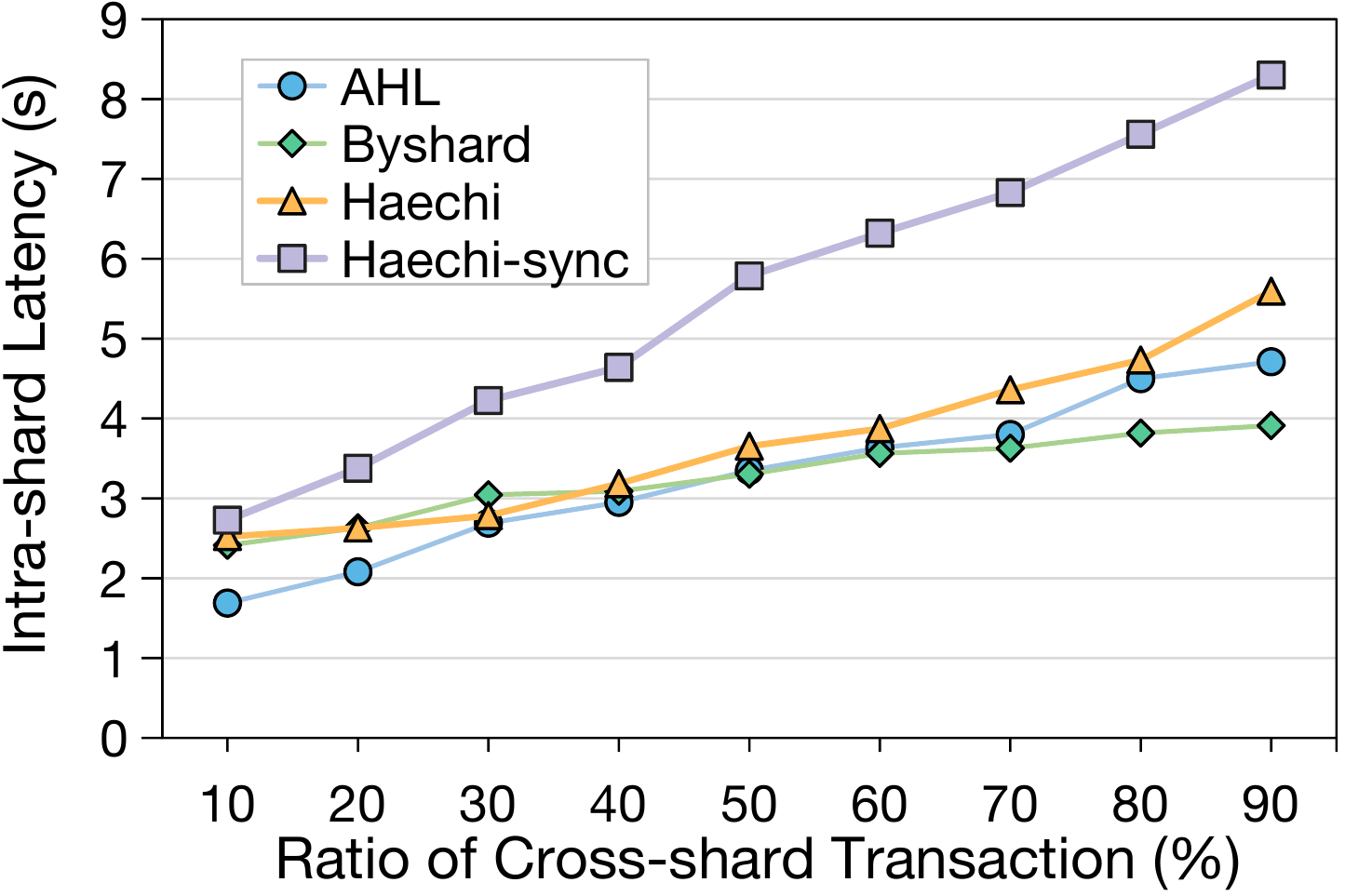}
		\caption{Intra-shard latency under varying ratios.}
		\label{fig-intra-latency-ratio}
	\end{minipage}
	\hfill
	\begin{minipage}[t]{0.23\textwidth}
		\centering
		\includegraphics[width=1.7in]{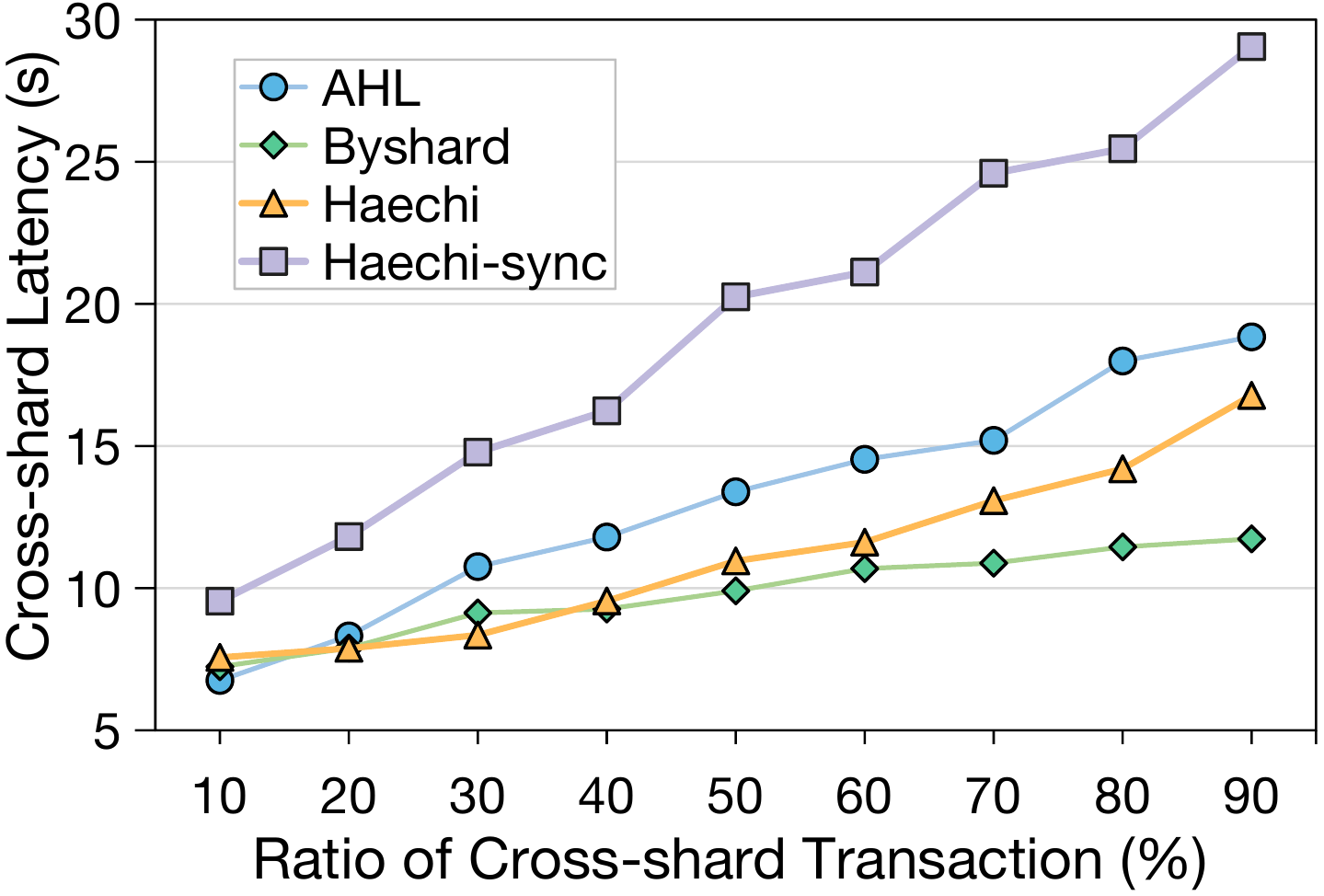}
		\caption{Cross-shard latency under varying ratios.}
		\label{fig-cross-latency-ratio}
	\end{minipage}
 % \vspace{-5mm}
\end{figure*}

\subsection{Micro-benchmark}
\label{section-security}
% To analyze the efficiency of our proposed fair ordering algorithm in \cref{section:haechi details-ordering phase}, we evaluate the generation interval of cross-shard call lists (CCLs) in the beacon chain. We run our protocol to wait for it to generate at least 30 CCLs, and compute the maximum, minimum, and average intervals for two consecutive CCLs. Note that the beacon chain will generate one CCLs once the CrossLinks it receives satisfy the conditions discussed in \cref{section:haechi details-ordering phase}. 

To evaluate the practicality of the global ordering phase introduced by both Haechi and Haechi-sync, we construct two experiments to calculate the CCLs interval and gas consumption respectively. The CCLs interval reflects the time spent in one ordering cycle, and the gas consumption measures how much extra gas will be used by the ordering phase.

First, we run both Haechi and Haechi-sync to wait for them to generate at least 10 CCLs, and compute the maximum, minimum, and average intervals for two consecutive CCLs. Fig. \ref{fig-shard-ccls} shows that the impact of the number of shards on CCLs generations. Due to the finalization fairness algorithm used in Haechi, the number of shards brings a subtle impact on CCLs generations, and the CCLs interval is acceptable under a real-world network latency, e.g., only 1.68s on average with 32 shards. In contrast, Haechi-sync introduces higher overhead into its ordering phase due to its synchronization mechanism.

\begin{table}[t]
\centering
\footnotesize
\caption{Evaluation of extra gas consumption of a Haechi-enabled sharded system with 32 shard chains.}
% (Gas price: 1 gas $\approx$ 15 Gwei = $1.5*10^{-8}$ Ether).}
\label{table:gas-consump}
\begin{threeparttable}
\begin{tabular}{|c|c|c|c|c|}
\hline
Protocol& $\#$CrossLink & MAX& AVG & MIN \\
\hline
Haechi-sync & 32 & 250 gas & 39 gas & 8 gas \\
Haechi& 10,000 & 78,000 gas & 103 gas & 8 gas \\
\hline
\end{tabular}
% \begin{tablenotes}
%     \footnotesize
%     \item[1] Assume at most 10,000 blocks are ordered in one ordering cycle.
% \end{tablenotes}
\end{threeparttable}
% \vspace{-6mm}
\end{table}

Then, to evaluate the practicality of our protocol on gas consumption, we convert the ordering phase into the number of comparisons of timestamps. Specifically, the ordering phase uses a sorting algorithm to order CrossLinks based on their timestamps, which involves a series of comparison operations. Therefore, we implement and deploy a comparison contract to estimate the gas consumption of one comparison for two timestamps, showing that one comparison operation only costs 780 gas. Besides, we assume the average number of transactions per CrossLink is 100 (note that transactions will amortize the gas), and at most 10,000 CrossLinks\footnote{10,000 means when receiving the first CrossLink from the slowest shard, the beacon chain has already received 9,999 CrossLinks from other shards, indicating $>300$ times difference in consensus speeds between shards.} are ordered in one ordering cycle of Haechi. Note that Haechi-sync (with 32 shards) always orders 32 CrossLinks in each ordering cycle. With these data, we can estimate the extra gas consumption needed in the ordering phase for each transaction. Table \ref{table:gas-consump} shows the upper bound of gas consumption, where MAX, AVG, and MIN are respectively corresponding to the complexities of a sorting algorithm in different cases, i.e., $O(n^2)$, $O(nlogn)$, and $O(n)$ times of comparisons with $n$ CrossLinks. From the historical gas price of Ethereum from July 2022 to July 2023\footnote{https://etherscan.io/chart/gasprice}, the maximum gas price is about 150 Gwei\footnote{1 Gwei=$10^{-9}$ ETH}. Therefore, the upper bound of gas consumption for each transaction in Haechi is worth $78,000*150$ Gwei = $1.17*10^{-2}$ ETH. Similarly, we can calculate the average fee in Haechi, $1.55*10^{-5}$ ETH, which is very cheap and negligible to the asset loss of users from a front-running attack.

% The experimental result in Fig. \ref{fig-shard-ccls} shows that the number of shards brings a subtle impact on CCLs generations, and the CCLs interval is acceptable under a real-world network latency, e.g., only 1.68s on average with 32 shards. Therefore, we can conclude that ordering transactions in the beacon chain won't bring heavy overhead and lead to a bottleneck with the increasing number of shards.

% \begin{figure}
% 	\setlength\abovecaptionskip{0.1\baselineskip}
% 	\setlength\belowcaptionskip{-1.0\baselineskip}   % 
% 	\centering
% 	\includegraphics[width=2.5in]{result/shard-ccls.pdf}
% 	\caption{The generation interval of CCLs.}
% 	\label{fig-shard-ccls}
% \end{figure}

\subsection{Performance}
\label{section-performance}
We next evaluate the performance of Haechi under different experimental settings. In these experiments, we deploy 33 shards (including 1 beacon shard) and use clients to continuously send transactions to derive the peak performance.  

Fig. \ref{fig-tps-intra-shard} and \ref{fig-tps-cross-shard} show the relationship between TPS and confirmation latency. We can find that Haechi only compromises slight performance decreasing compared to Byshard, and even performs better than AHL; while Haechi-sync sacrifices marked performance loss. Haechi has less than 5s confirmation latency for intra-shard transactions and less than 10s confirmation latency for cross-shard transactions when the system has a TPS of less than 15,000, which can satisfy most requirements of decentralized applications in a practical scenario.

Fig. \ref{fig-tps-client}, \ref{fig-intra-latency-client}, and \ref{fig-cross-latency-client} present the performance of different cross-shard protocols under distinct workloads, in which we respectively instantiate $c=10, 50, 100, 1000, 2000, 3000$ clients for a shard to establish connections with the shard nodes and send transactions to them. These experiments are used to simulate a real-world scenario where users concurrently send requests to the system. The results show that in our experimental environment, all cross-shard protocols achieve the peak TPS when the number of clients in a shard exceeds 2,000. But the confirmation latency will still increase with more clients. This is because 
transactions need to wait more time in the mempool once the speed of transaction requests is larger than the speed of handling transactions by the system. Besides, we observe committing a new block ranges from 1s to 8s in different shards because of region delays and transaction workloads. This shows the existence of the consensus difference in a realistic sharded system.
% Tendermint needs resources to establish connections for each request and too many requests will cause the system to fail to receive transactions in time, prolonging the confirmation latency. Such a design might lead to a DDoS attack. But we could adopt some mechanisms to prevent it, e.g., the blacklist mechanism. We leave this implementation to our future work.

Fig. \ref{fig-tps-ratio}, \ref{fig-intra-latency-ratio}, and \ref{fig-cross-latency-ratio} give the impact of the ratio of cross-shard transactions on the performance, where we set this ratio from $r=10\%$ to $90\%$. For example, $r=90\%$ represents that an average of 9 out of 10 transactions are cross-shard transactions. From the experimental results, we can find that with an increasing ratio of cross-shard transactions, the performance of a sharded system will decrease. This is because handling cross-shard transactions will introduce many overheads in communication and computation, which reflects the significance of designing an efficient cross-shard protocol. We hope our designed asynchronous and non-blocking processing for cross-shard transactions (\S~\ref{section5-verification}) can motivate readers to explore more efficient cross-shard protocols.

From the above evaluations, we can find Haechi only sacrifices a slight performance decrease compared to the 2P cross-shard consensus protocol while Haechi also provides finalization fairness property for a sharded system.

%% file: appendics/apdx-relatedworks.tex
\section{Related Work}\label{related-work}

\noindent \textbf{Blockchain sharding.} Sharding has been proposed to enhance blockchain scalability. 
A major new challenge is how to handle cross-shard transactions. Many works \cite{Monoxide,rivet} adopt the optimistic commit protocol to guarantee isolation and atomicity, where cross-shard transactions are divided into several sub-transactions to related shards and committed optimistically. However, this protocol needs to abort and roll back all transactions if there are conflicting read/write of the same data. Another cross-shard protocol is the two-phase (2P) protocol as we discussed above, which is adopted by most works \cite{chainspace,Omniledger,Rapidchain,ahl, skychain, sharper, byshard}. 

Due to the ledger separation, handling a cross-shard transaction inevitably involves several steps. In each step, a related shard runs one instance of intra-shard consensus to handle related data. Thus, compared to intra-shard transactions, handling cross-shard transactions with the previous cross-shard consensus protocol introduces several instances of the intra-shard consensus. To the best of our knowledge, none of the previous works care about the differences between intra-shard and cross-shard consensus protocols. However, as we discussed above, these differences will introduce the finalization fairness problem to a sharded system and bring adverse effects that will further limit the application of sharding technology. Instead, Haechi can address the finalization fairness problem without compromising too much performance.

\noindent \textbf{Order-fairness Protocols.} With the popularity of DeFi, a new property called \emph{order fairness} is gaining more and more attention. Many recent works have shown that adversaries can make profits by manipulating transaction order \cite{eskandari2019sok,flashboys,sandwich-attack,front-running-discovery,quantify-front-attack,zhou2022sok}. Accordingly, various fair consensus protocols have been proposed to achieve order fairness for blockchains. Some of them \cite{byzantine-oligarchy, wendy, aequitas,themis,cachin2021quick} adopt the \emph{time-based} order-fairness, where transactions are ordered based on the time when they are received by nodes. The others \cite{asayag2018fair,stathakopoulou2021adding, malkhi2022maximal}
adopt the \emph{blind order-fairness}, where the content of transactions is hidden until they are committed to a total ordering, and thus transaction order is random. 

All the above works consider how to form order fairness during the period from when transactions appear in the mempool to when transactions are packed into blocks. However, Haechi aims to achieve finalization fairness in a sharded system, focusing on order fairness during the period from when transactions are first processed in the sender shards to when transactions are executed in the contract shards. Besides, most of them assume a permissioned blockchain scenario, which only allows a limited number of consensus nodes. In contrast, Haechi provides a new solution for achieving order fairness in permissionless blockchains. 

%% file: discussion.tex
\section{Discussion}\label{sec-discussion}
\noindent \textbf{Block timestamp security.} In \S~\ref{section:haechi details-ordering phase}, we rely on the correct block timestamp to achieve finalization fairness. While all prominent blockchain designs ensure that block proposers cannot set arbitrary timestamps \cite{aptos-timestamp, eth-timestamp}, the proposers still have small leverage to modify the timestamp within tolerance introduced by partial synchrony or asynchrony so that they can influence the execution order. Nevertheless, such a minor-level timestamp manipulation seems unavoidable across fairness protocols that use timestamps as the ordering indicator even with synchronized clocks \cite{byzantine-oligarchy, aequitas, cachin2021quick}. A mechanism to mitigate this risk is to hide transaction information using an encryption scheme until its block reaches a consensus. We denote $T_c$ as the time used for this consensus process. In this case, when the front-running proposer observes the content of the transaction, it needs to attach a timestamp with a time deviation of larger than $T_c$ compared to the actual time. However, such a marked deviation of a block timestamp could lead to rejection on this block by honest nodes.

\noindent
\textbf{Beacon chain-free Haechi.} Currently, Haechi assigns the beacon chain to coordinate/lead the ordering phase and establish a globally fair order via our finalization fairness algorithm. However, we emphasize our algorithm can be extended to any sharded system even if it has not the beacon chain. This will formulate a so-called \emph{beacon chain-free Haechi} or \emph{leaderless Haechi}. Specifically, instead of sending CrossLinks to the specific beacon chain, all shards need to broadcast their CrossLinks to other shards. Then each shard can run the finalization fairness algorithm locally to establish a global order. Note that our algorithm %can 
establishes a \emph{deterministic} global order as we use the block timestamps as the ordering indicator and the block timestamps are globally consistent, i.e., for any node, the timestamp of a block is consistent. With the deterministic guarantee, all shards can eventually achieve a consistent view of the global order. However, this all-to-all communication brings significantly more communication overhead as compared to the current (all-to-one and then one-to-all) beacon/leader chain-based approach.
% the beacon chain-free Haechi is more general and more resilient to a single point of performance bottleneck. We leave its implementation to our future work.

% \bigbreak
\noindent \textbf{Intra-shard order fairness.} This work focuses on the finalization fairness of sharded systems, i.e., the execution order of transactions is the same as their processing order. More precisely, we guarantee \emph{cross-shard order fairness} against the front-running attack presented in this paper. However, the adversary can manipulate the transaction order in one shard chain, breaking \emph{intra-shard order fairness}. Specifically, the adversary controls the intra-shard consensus of the shard to order their transactions before the victim's transactions in an adversary shard block. In this case, the adversary manipulates the processing order, thus affecting the finalization fairness.

Nevertheless, the order fairness problem for intra-shard consensus can be mitigated by existing order-fairness protocols as discussed in \S~\ref{related-work}. Haechi focuses on finalization fairness across shards, we omit the intra-shard order fairness from our evaluations. However, we emphasize that these order-fairness protocols are complementary and compatible with Haechi and can be easily integrated into Haechi. % as the intra-shard consensus protocol to make a sharded system absolutely fair.

% \bigbreak
\noindent \textbf{Multi-shard transaction order.} In a real-world scenario, a transaction can involve multiple smart contracts by \emph{contract interactions}, where some contract operations, e.g., function-call and data-return, will invoke other contracts. Such a transaction is also called \emph{multi-shard contract transaction} in a sharded system if its involved contracts are managed by multiple shards. A multi-shard transaction will be handled by more than two shards and may involve at least two block timestamps until being executed in the OSC. Haechi can still ensure finalization fairness by using the \emph{first block timestamp} when the multi-shard transaction is first packed into a shard block. In the implementation, the beacon chain maintains a mapping data structure to trace the first block timestamp for multi-shard transactions so that it can ensure the execution order is consistent with the processing order (i.e., ensure finalization fairness). There are some potential optimizations for it; however, they are beyond the scope of this paper.

\noindent \textbf{Impact of reconfiguration on the attack.} The proposed front-running attack relies on the attacker observing the victim's transactions in time when they are in the same shard. It is part of the natural consensus flow of communication across shards. While the attack is in process, if the attacker is swapped to another shard via reconfiguration, it may not be able to successfully complete the attack. However, if nodes are reorganized so frequently, then a system may not be able to process transactions either, which is impractical. For instance, the reconfiguration interval (called \textit{epoch}) for Ethereum is about $384$ seconds as each epoch contains $32$ blocks. Therefore, attackers still have the advantage of launching front-running attacks in their shards. Reconfiguration cannot compromise the effectiveness of the front-running attack without reducing the system throughput to near zero.

%% file: conclusion.tex
\section{Conclusion}\label{section: conclusion}
In this paper, we explore the finalization fairness problem in sharded systems. We present, to the best of our knowledge, the first front-running attack targeting the existing cross-shard consensus protocols. Then, we propose Haechi, a new cross-shard consensus protocol that provides finalization fairness for users from different shards. Haechi introduces a global ordering phase to prevent attackers from intentionally manipulating the execution order of transactions calling those smart contracts that are vulnerable to front-running attacks in sharded systems. We implement a full prototype and compare Haechi to other cross-shard consensus protocols. Our results show that Haechi can achieve approximate throughput and confirmation latency as other cross-shard protocols (i.e., 13,000+ TPS and ~17s cross-shard latency with 33 shards). We also evaluate the presented cross-shard front-running attacks. Experimental results show the effectiveness of the attacks on existing cross-shard protocols while Haechi can prevent such attacks well.

%% file: appendix.tex
% \section{Appendix}
% \label{sec-appendix}

% \input{appendics/apdx-relatedworks}
% \input{appendics/apdx-proof-v2}
\input{appendics/apdx-attackmodel}
\input{appendics/apdx-blockprocess}

%% file: appendics/apdx-attackmodel.tex
\section{A more flexible Front-running Attacking Model in Sharded Systems}\label{appendix-general-attack}
In \cref{section-attack-process}, we give a concrete example that the adversary launches a front-running attack by utilizing the processing-execution difference between intra-shard and cross-shard transactions. Briefly, the adversary creates intra-shard transactions to directly call a contract via intra-shard consensus, once it monitors that the victims' cross-shard transactions from other shards are about to call the same contract. We call this attacking model triggered by intra-shard transactions as the \emph{intra-shard attacking model}.

Except for the processing-execution difference between intra-shard and cross-shard transactions, there also exists a \emph{processing-execution difference between cross-shard transactions from distinct shards}. To be more specific, shards have distinct consensus speeds and network transmission capabilities because of resource diversity, leading that transactions are processed and transferred across shards at various speeds. The resource diversity in a real-world scenario can include computing power, shard size, network bandwidth, etc. In this section, we will discuss a more flexible front-running attacking model based on such a difference in a sharded system, called \emph{cross-shard attacking model}.

\noindent 
\textbf{Cross-shard attacking model.} The adversary utilizes the processing-execution difference between cross-shard transactions from distinct shards to launch a front-running attack. Specifically, the adversary can gather in a shard that has the fastest consensus speed (i.e., the shortest block interval) and the highest bandwidth (i.e., the best network connection). By doing this, the attacker's cross-shard transactions will suffer a shorter processing time and delivery time, thus front-running to be executed by the contract's shard compared to the victims' cross-shard transactions. 

\begin{figure}[t]
	\setlength\abovecaptionskip{-0.05\baselineskip}
	\setlength\belowcaptionskip{-1.2\baselineskip}   % 
	\centering
	\includegraphics[width=3.4in]{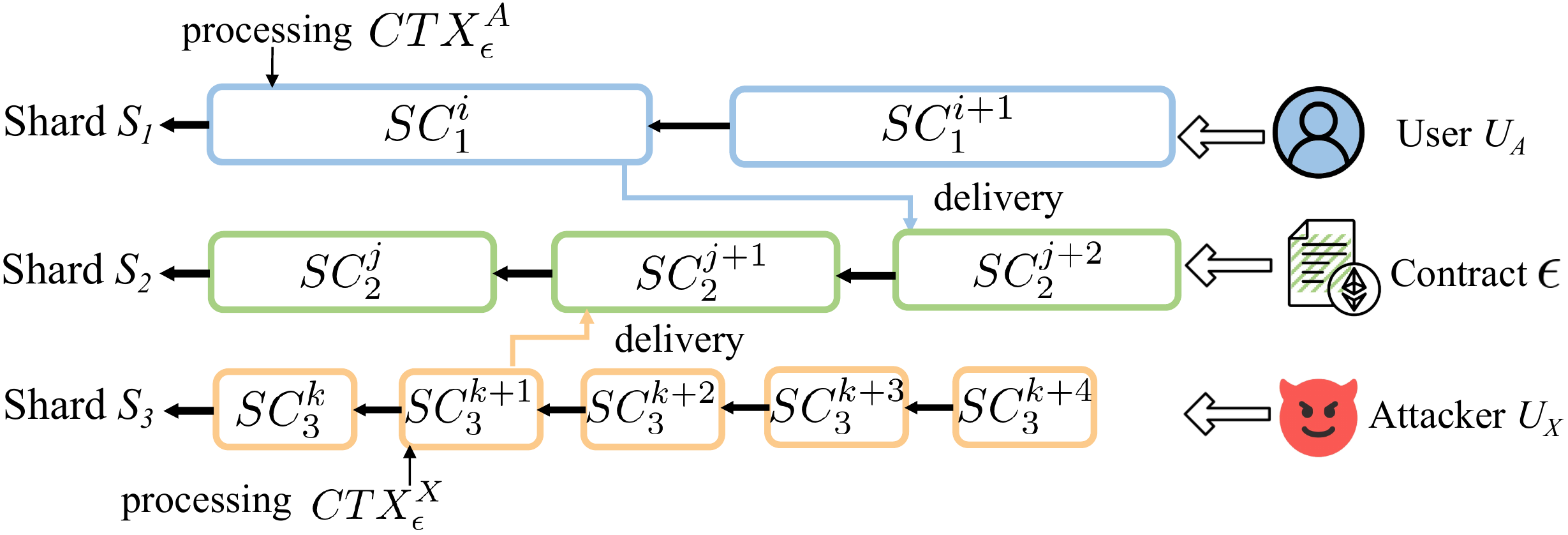}
	\caption{The cross-shard attacking: the adversary registers accounts in a shard with faster consensus speed, where it utilizes the processing-execution difference between cross-shard transactions to launch a front-running attack.}
	\label{fig-mediate-cross-attacking}
\vspace{-0.1cm}
\end{figure}

\begin{figure*}
    \setlength\abovecaptionskip{-0.02\baselineskip}
    \setlength\belowcaptionskip{-1.4\baselineskip} 
    \centering
    \includegraphics[width=6.5in]{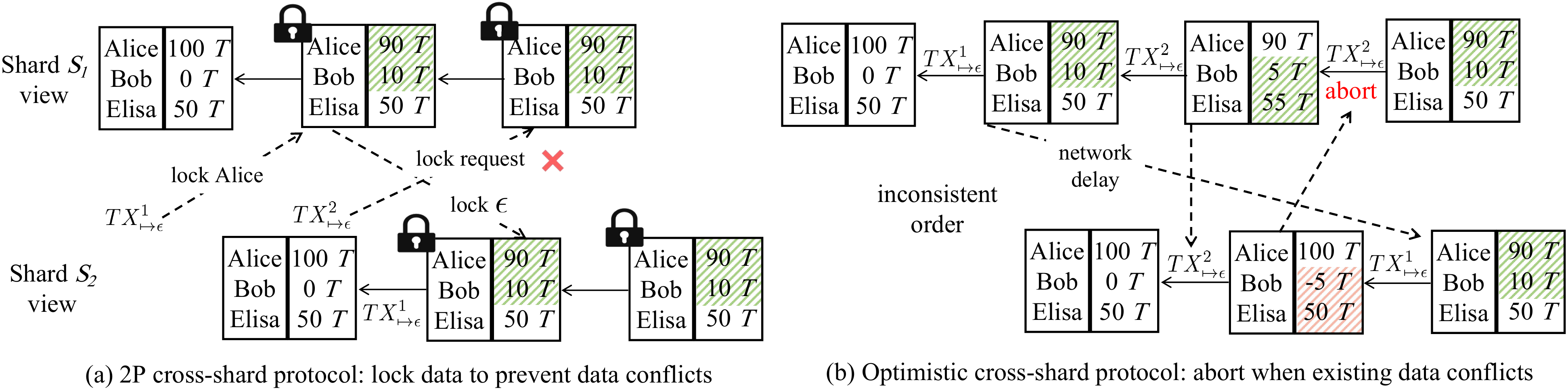}
    \caption{Different concurrency control mechanisms for handling cross-shard transactions.}
    \label{fig-process-way}
    \vspace{-0.1cm}
\end{figure*}

Fig. \ref{fig-mediate-cross-attacking} shows a concrete example of the cross-shard attacking model. We use the rectangle to represent a shard block and its length represents the consensus speed of this shard where a longer length means a longer time needed for consensus. Besides, from left to right of this figure is the increasing timestamp. We assume in this figure, shard $S_1$ maintains a victim's account $U_A$, $S_2$ maintains an order-sensitive contract $\epsilon$, and $S_3$ maintains the attacker's account $U_X$. As shown in Fig. \ref{fig-mediate-cross-attacking}, $S_3$ can process transactions faster than $S_1$. In this case, once the attacker $U_X$ monitors that a victim's cross-shard transaction (e.g., $CTX^A_{\epsilon}$) is about to call $\epsilon$, it immediately creates a cross-shard transaction (e.g., $CTX^X_{\epsilon}$) in $S_3$. Since $S_3$ has faster consensus speed and shorter delivery network latency than $S_1$, the front-running transaction $CTX^X_{\epsilon}$ can be executed by $S_2$ even though $CTX^A_{\epsilon}$ is processed earlier (i.e., in block $SC^i_1$) than $CTX^X_{\epsilon}$ (i.e., in $SC^{k+1}_3$). Similarly, such a front-running attack can also happen to \emph{every shard} that has a slower consensus speed or longer network latency than $S_3$.

Compared to the intra-shard attacking model, the adversary won't need to register multiple accounts in different contracts' shards with the cross-shard attacking model. For example, in the scenario of Fig. \ref{fig-mediate-cross-attacking}, the intra-shard attacking model asks the adversary to respectively register accounts in $S_2$ and other contracts' shards so that the adversary can construct intra-shard transactions to front-run the victim's cross-shard transactions. On the contrary, the adversary with the cross-shard attacking model only needs to register an account in the shard $S_3$ that has faster consensus speed and higher network bandwidth than other shards.

\noindent \textbf{Haechi is immune to these attacking models.} We find that these front-running attacking models rely on the processing-execution differences between intra-shard and cross-shard transactions or between cross-shard transactions from different shards. However, in Haechi, such processing-execution differences are eliminated since transactions calling the same contract are globally ordered before being executed via our finalization fairness algorithm proposed in \cref{section:haechi details-ordering phase}. Therefore, the execution order can be guaranteed consistent with the processing order, and Haechi ensures the finalization fairness.

%% file: appendics/apdx-blockprocess.tex
\section{Blocking and Non-blocking Processing}\label{appendix-blocking-verification}
% \textbf{Verification dilemma.} 
Sharding divides nodes into multiple shards to allow them to handle transactions in parallel. For intra-shard transactions from different shards, they can be processed and committed in parallel since they have \emph{no data conflict}. However, for cross-shard transactions that may access the same data, a sharded system requires a concurrency control mechanism, i.e., cross-shard protocol, to ensure isolation and atomicity. For illustration, we assume two conflicting cross-shard transactions: (i) $TX^1_{\mapsto \epsilon}$: Alice in shard $S_1$ calls a smart contract $\epsilon$ in shard $S_2$ to transfer its 10 Tokens (denoted $T$) to Bob; (ii) $TX^2_{\mapsto \epsilon}$: Bob in shard $S_1$ calls the smart contract $\epsilon$ in shard $S_2$ to transfer its 5 $T$ to Elisa. Note that Alice and Bob have their accounts maintained by $S_1$ while the token balances of Alice, Bob, and Elisa are a part of the contract state of $\epsilon$ maintained by $S_2$. Furthermore, we assume that Alice, Bob, and Elisa respectively have 100 $T$, 0 $T$, and 50 $T$ at the beginning. Fig. \ref{fig-process-way} illustrates two concurrency control mechanisms for handling these two conflicting cross-shard transactions: 2P cross-shard protocol and optimistic cross-shard protocol.

% \bigbreak
\noindent \textbf{Blocking processing.} The 2P cross-shard protocol is a blocking processing mechanism where the processing of a later conflicting cross-shard transaction is blocked until the end of the cross-shard consensus of a former cross-shard transaction. The main idea of the 2P cross-shard protocol is to lock all relevant data to ensure isolation and atomicity when handling cross-shard transactions in parallel. It works well when two cross-shard transactions have no data conflicts, i.e., they don't have access to the same data and can be processed concurrently. However, it will block the processing of later conflicting cross-shard transactions if the previous cross-shard transaction keeps locking the conflicting data. As shown in Fig. \ref{fig-process-way}(a), when $S_1$ and $S_2$ coordinately handle $TX^1_{\mapsto \epsilon}$, the contract state of $\epsilon$ is locked. If the conflicting $TX^2_{\mapsto \epsilon}$ hopes to request the lock to call $\epsilon$, it will fail until the lock of $\epsilon$ is released by $TX^1_{\mapsto \epsilon}$. In this case, these two conflicting cross-shard transactions have to be handled one by one, i.e., blocking processing. Obviously, the blocking processing limits the capability of a sharded system in handling conflict cross-shard transactions. Especially in a real-world blockchain network, there are some popular smart contracts that are called frequently by users' transactions, and it is challenging to process these conflicting transactions well with the blocking processing mechanism.

% \bigbreak
\noindent \textbf{Non-blocking processing.} The optimistic cross-shard protocol can achieve non-blocking processing for cross-shard transactions. However, this protocol will introduce a high abortion rate when there are numerous conflicting transactions. As shown in Fig. \ref{fig-process-way}(b), $S_1$ first handles $TX^1_{\mapsto \epsilon}$, and it \emph{optimistically} assumes $TX^1_{\mapsto \epsilon}$ will be executed successfully, after which Bob will have 10 $T$ in the contract $\epsilon$. Then, $S_1$ starts processing $TX^2_{\mapsto \epsilon}$ without waiting for the final commitment of $TX^1_{\mapsto \epsilon}$. However, a network delay leads that $TX^2_{\mapsto \epsilon}$ arrives at $S_2$ before $TX^1_{\mapsto \epsilon}$. In this case, $TX^2_{\mapsto \epsilon}$ will be executed by $S_2$ before $TX^1_{\mapsto \epsilon}$. However, since the token balance of Bob is 0 $T$ at that time, $TX^2_{\mapsto \epsilon}$ will lead to a negative value and thus fail to be executed. Finally, $TX^2_{\mapsto \epsilon}$ is aborted in both $S_1$ and $S_2$.

Comparing the two concurrency control mechanisms, we find that the optimistic cross-shard protocol is more suitable to be used in a practical scenario. Because there are always some popular smart contracts (e.g., DEXs discussed in \S~\ref{section-attack-impact}) in a real-world blockchain network. The non-blocking processing property enables a sharded system to handle more transactions that call these popular smart contracts over time, without introducing a long-time lock waiting. However, the existing optimistic cross-shard protocol introduces frequent abortions if these transactions have numerous read/write conflicts for the same popular contracts. The reason why the optimistic cross-shard protocol leads to transaction abortions is that \emph{conflicting transactions are out-of-order} before they are executed. To be more specific, since there is no global order for transactions that call the same contract, relevant shards may have different orders when processing these transactions. These inconsistent processing orders break the consistency and isolation of transactions and thus lead to transaction abortions.

% \bigbreak
\noindent \textbf{Non-blocking processing in Haechi.}  Different from the existing optimistic cross-shard protocols, Haechi introduces an ordering phase to establish a global order of transactions for \emph{each OSC}. By doing this, transactions from different shards calling the same OSC will have a \emph{deterministic sequential order}. The deterministic order eliminates the concurrency conflicts and thus allows shards to independently and non-blocking process transactions. For those non-conflict contracts that do not interact with each other, Haechi can still handle their contract transactions in parallel. Therefore, compared to previous cross-shard protocols, Haechi can better process conflicting transactions in a non-blocking way and still support handling non-conflict transactions in parallel. This is promising to apply in a practical blockchain scenario.